\documentclass[pdflatex,sn-mathphys-num]{sn-jnl}

\usepackage{graphicx}%
\usepackage{multirow}%
\usepackage{amsmath,amssymb,amsfonts}%
\usepackage{amsthm}%
\usepackage{mathrsfs}%
\usepackage[title]{appendix}%
\usepackage{xcolor}%
\usepackage{textcomp}%
\usepackage{manyfoot}%
\usepackage{booktabs}%
\usepackage{algorithm}%
\usepackage{algorithmicx}%
\usepackage{algpseudocode}%
\usepackage{listings}%
\usepackage{caption}

\theoremstyle{thmstyleone}%
\theoremstyle{thmstyletwo}%

\theoremstyle{thmstylethree}%

\def\xbI{{J161201}}
\def\xbIII{{J061331}}
\newcommand\boldt{}  

\begin{document}

\title[Article Title]{XMM-\textit{Newton} follow-up of two eROSITA X-ray binary candidates\textsuperscript{$\bigstar$}}

\begingroup
\renewcommand\thefootnote{}\footnotetext{\textsuperscript{$\bigstar$} Partly based on observations made with the Southern African Large Telescope (SALT).}
\addtocounter{footnote}{-1}
\endgroup


\author*[1]{\fnm{A.} \sur{Avakyan}}\email{artur.avakyan@astro.uni-tuebingen.de}

\author[2]{\fnm{A,} \sur{Zainab}}\email{aafiazainab.ansar@fau.de}

\author[1]{\fnm{V.} \sur{Doroshenko}}\email{doroshv@astro.uni-tuebingen.de}

\author[2]{\fnm{J.} \sur{Wilms}}\email{joern.wilms@sternwarte.uni-erlangen.de}

\author[3]{\fnm{A.} \sur{Schwope}}\email{aschwope@aip.de}

\author[1]{\fnm{V.} \sur{Suleimanov}}\email{suleimanov@astro.uni-tuebingen.de}

\author[4]{\fnm{D.} \sur{Buckley}}\email{dibnob@saao.ac.za}

\author[3]{\fnm{J.} \sur{Brink}}\email{jbrink@aip.de}

\author[1]{\fnm{A.} \sur{Santangelo}}\email{santangelo@astro.uni-tuebingen.de}

\affil*[1]{\orgdiv{Institut f{\"u}r Astronomie und Astrophysik T{\"u}bingen}, \orgname{Universit{\"a}t T{\"u}bingen}, \orgaddress{\street{Sand 1}, \city{T{\"u}bingen}, \postcode{72076}, \state{Baden-W{\"u}rttemberg}, \country{Germany}}}

\affil[2]{\orgdiv{Dr. Karl-Remeis Sternwarte and Erlangen Centre for Astroparticle Physics}, \orgname{Friedrich-Alexander Universit{\"a}t Erlangen-N{\"u}rnberg}, \orgaddress{\street{Sternwartstr. 7}, \city{Bamberg}, \postcode{96049}, \state{Bavaria}, \country{Germany}}}

\affil[3]{\orgname{Leibniz-Institut f{\"u}r Astrophysik Potsdam}, \orgaddress{\street{Sternwarte 16}, \city{Potsdam}, \postcode{14482}, \state{Brandenburg}, \country{Germany}}}

\affil[4]{\orgdiv{Department of Astronomy}, \orgname{University of Cape Town}, \orgaddress{\street{Private Bag X3}, \city{Rondebosch}, \postcode{7701}, \country{South Africa}}}


\abstract{We report on the follow-up observations with XMM-\textit{Newton} of two X-ray binary candidates identified in the first eROSITA all-sky survey data (eRASS1), 1eRASS\,J061330.8$+$160440 and 1eRASS\,J161201.9$-$464622. Based on the obtained results, in particular, the observed X-ray spectra and lack of pulsations, as well as properties of the identified optical counterparts, we conclude that both candidates are \boldt{unlikely to be XRBs. Based on LAMOST optical spectroscopy and SED fit results for 1eRASS\,J061330.8$+$160440 we classify it as an M0 chromospherically active subgiant star. ZTF and TESS photometry reveal highly significant period for this object of 7.189 days, which likely attributed to starspot(s). On the other hand, SALT follow-up spectroscopy of 1eRASS\,J161201.9$-$464622 solidly classifies this source as a bright novalike cataclysmic variable (CV), the second discovered with eROSITA. A persistent 4.802\,h signal is found across all three available TESS observations, and is tentatively identified as the orbital period of the binary. Follow-up high-speed photometry and time-resolved spectroscopy are required to confirm the derived orbital modulation.}}


\maketitle

\section{Introduction}\label{sec1}

Searching for faint X-ray binaries (XRBs) in all-sky surveys requires accurate identification of their counterparts in other bands and detailed analysis of their multi-wavelength (MWL) properties. However, since most XRBs in the Galaxy are expected to have relatively low X-ray luminosities \citep{Doroshenko14pop}, observations of faint XRBs are essential to understand the properties of the XRB population as a whole, and to study accretion physics at low luminosities. 
Up to now, however, only a handful of such faint XRBs are known~\citep{Avakyan23, Marvin23, Fotin}\footnote{\url{http://astro.uni-tuebingen.de/~xrbcat/}}$^{,}$\footnote{\url{https://binary-revolution.github.io/}}. In particular, the majority of persistently long-term faint XRBs have been discovered in observations of other targets \boldt{or in deep inspection of the Galactic centre or globular clusters}~\citep{2003ApJ...598..501H,2003ApJ...589..225M,2016ApJ...825...10T,2017MNRAS.467.2199B,2021ApJ...921..148M}. This was due to the low sensitivity of hard X-ray surveys and strong absorption in the Galactic plane (where most XRBs are located).

The situation, however, began to change thanks to the all-sky survey performed by the extended ROentgen Survey Imaging Telescope Array \citep[eROSITA,][]{Merloni23, Merloni12, eROSITA} on board the Spectrum-Roentgen-Gamma (Spektra-RG, SRG) mission. Indeed, eROSITA offers much higher sensitivity compared to previously available all-sky surveys in a similar energy band \citep[by a factor 20–40 compared to Roentgensatellite (ROSAT) All-Sky Survey,][]{ROSAT_miss, ROSAT99, ROSAT} or compared to hard X-ray observatories such as the INTErnational Gamma-Ray Astrophysics Laboratory \citep[INTEGRAL,][]{Integral}. In addition, it observes in a significantly wider energy band (0.2-10\,keV compared to the 0.2-2\,keV band covered by ROSAT), which reduces the effects of the absorption in the Galactic plane. eROSITA provides X-ray spectral information, variability on several timescales, and significantly improved point source localisation accuracy compared to all aforementioned wide-field X-ray surveys.

Of the four all-sky surveys completed by eROSITA (eRASS1-4, with a duration of six months each) as part of its deep X-ray all-sky exploration. As part of the German eROSITA Consortium, we have data rights to half the sky (western Galactic hemisphere), of which the first survey has recently been released for public usage~\citep{Merloni23}. 
A majority of the X-ray sources detected by eROSITA are active galactic nuclei (AGNs), ordinary stars, and galaxy clusters (GCs)~\citep{Merloni12, eROSITA}. 
However, we also expect to discover a large number~(up to a few hundreds) of Galactic X-ray binaries, which will help obtain important constraints on their X-ray luminosity function~\citep{Lutovinov13, Doroshenko14pop}. These constraints mainly concern the low luminosity part of the function, where eROSITA is significantly more sensitive compared to other wide-field instruments.

Identification of faint XRBs among millions of other sources is, however, challenging due to the similarity of their X-ray properties to those of AGNs and stars dominating the survey. Efforts aimed at the classification of all sources detected by eROSITA are ongoing and will follow the first data release~(eRASS\,DR1). In the meantime, several XRB candidates were identified based on the preliminary analysis and classification of X-ray sources detected in the first eROSITA all-sky survey (eRASS1). In particular, the combination of eROSITA X-ray fluxes and near/mid infrared properties from Two Micron All Sky Survey~\citep[2MASS,][]{2MASS_miss, 2MASS, 2MASS_1} and All-Sky Wide-field Infrared Survey Explorer~\citep[AllWISE, WISE,][]{WISE, WISE_cat, CW2020} has been used to identify XRB candidates for follow-up observations with the X-ray Multi-Mirror Mission~\citep[XMM-\textit{Newton},][]{XMM_mission} and the Nuclear Spectroscopic Telescope ARray~\citep[NuSTAR,][]{NuStar}. In addition, in order to better identify candidates, three new catalogues of high-mass~(HMXB) and low-mass~(LMXB) X-ray binaries by \cite{Marvin23,Fotin} and \cite{Avakyan23}, respectively, have been used.

Four XRB candidates proposed for X-ray follow-up have been actually observed by this point, the two brighter ones by NuSTAR and the remaining two by XMM-\textit{Newton}. The detailed follow-up results of the first of the two objects followed up by NuSTAR, SRGA\,J124404.1$-$632232 has been presented in \cite{Doroshenko22Puls}. The source was classified to be a HMXB consisting of an accreting pulsar with a pulsation period $\sim538$\,s and a Be-type star as the optical counterpart, making it the first Galactic BeXRB discovered by eROSITA. Analysis of NuSTAR observations of the second candidate, 1eRASS\,J085039.9$-$421151, which was found to be a HMXB/Symbotic XRB with a likely neutron star compact object is presented in \cite{2024arXiv241102655Z}.

In this work, we report on the analysis of XMM-\textit{Newton} and eROSITA observations of the two XRB candidates (1eRASS$\,$J061330.8$+$160440 and 1eRASS\,J161201.9$-$464622, hereinafter \xbIII{} and \xbI{}). The structure of the paper is as follows: in Sect.~\ref{data_reduc} we first briefly summarise the available eROSITA and XMM-\textit{Newton} data. Then, in the same section, we describe the identification of plausible optical counterparts and their properties.
\boldt{In Sect.~\ref{data_anal}, we give a detailed spectral and timing analysis of the X-ray properties of the candidates. The same section also presents both spectroscopic and photometric analyses of the two candidates, along with the results of spectral energy distribution fitting for \xbIII{}.}
Finally, we discuss the likely origin of both sources and summarise the results in Sect.~\ref{disc} and Sect.~\ref{conc}, respectively.

\begin{table}[t]
\renewcommand{\arraystretch}{1.4} 
\renewcommand{\tabcolsep}{2mm}   
    \centering
    \caption{Summary of eROSITA observations of both sources.}
    \begin{tabular}{@{}lllll@{}}
    \toprule
     Source & eRASS &  $T_{\rm start}$, MJD  & $T_{\rm stop}$, MJD & Exposure, ks \\
     \midrule
        \xbIII{} &  1 & 58943.25  & 58944.08 & 0.147
        \\
        & 2 & 59129.13  &  59131.00 & 0.123
        \\
        & 3 & 59305.46  & 59306.62 & 0.203  
        \\
        & 4 & 59489.38 & 59490.71 & 0.216 
        \\
        \xbI{} & 1 & 58918.34  & 58920.01 & 0.260
        \\
        & 2 & 59101.45  & 59103.28 &  0.274 \\
        & 3 & 59272.07  & 59282.22 & 0.258
         \\
        & 4 & 59460.03  & 59461.86 & 0.284
        \\
    \botrule
    \end{tabular}
    \label{er_tabl}
\end{table}

\section{Observations and MWL counterparts}\label{data_reduc}
\subsection{X-ray observations and basic data reduction}\label{erview}

The initial selection and classification of both sources studied here was based on the preliminary analysis of eRASS1 data using the reduction pipeline available at the time (c946). Since then, three more surveys were carried out (eRASS2, eRASS3, eRASS4) and several updates to the source detection pipeline were implemented. We use the current (c020) eROSITA data analysis software eSASS~\citep{Brunner22} pipeline configuration (software calibration file version 211214) throughout the paper. Details of the observation times are presented in the Table~\ref{er_tabl}. The details of XMM-\textit{Newton} observations are summarised in Table~\ref{xmm_tabl}. We use the full data-set for characterising the X-ray properties of the candidates, but emphasise, that candidate selection was based on the first survey data only and used an earlier version of data processing.

The eROSITA source data products (events, light curves and spectra) were extracted using the eSASS commands \texttt{evtool} and \texttt{srctool}. Source fluxes were then determined based on modelling of the obtained spectra as described in Sect.~\ref{spectra}). 
The XMM-\textit{Newton} data reduction was carried out using XMM-\textit{Newton} analysis software XMM SAS 13.5 package, current calibration files and standard filtering criteria\footnote{\url{https://www.cosmos.esa.int/web/xmm-newton/sas-threads/}}. Cleaned event list files were produced by commands \texttt{evselect} and \texttt{tabgtigen} which were then used to select source and background photons from source-centred circular and annular extraction regions using the \texttt{xmmselect} task. In particular, we used annuli with radii of ${\sim}65''$ and ${\sim}132''$ (${\sim}77''$ and ${\sim}155''$) for background, and a source circle with radius of ${\sim}22''$ (${\sim}19''$) for \xbIII{} (\xbI{}). Radii for source selection region were optimised using the \texttt{eregionanalyse} task. Source detection was carried out with \texttt{emldetect} routine to derive final X-ray positions for both sources after applying astrometric correction as detailed in Sect.~\ref{x_ray_pos}. 
The source light curves and spectra were then extracted using the \texttt{lcurve} and \texttt{especget} tasks. All spectra were  grouped using the \texttt{specgroup} task to contain at least one count per energy bin to enable use of Cash statistics \citep{Cash79}. 
Finally, we have also extracted source photons in the energy range $0.5-10$\,keV using the same extraction regions to search for possible pulsations. All timing products were corrected to Solar system barycentre via \texttt{barycen} command. Spectral and timing analysis of the extracted high-level products was then finally carried out with the help of the High Energy Astrophysics Software HEASOFT~v6.29\footnote{\url{https://heasarc.gsfc.nasa.gov/docs/software/heasoft/}}.

\begin{table}[t]
\renewcommand{\arraystretch}{1.4} 
\renewcommand{\tabcolsep}{2mm}   
    \centering
    \caption{Summary of XMM-\textit{Newton} observations of both sources.}        
    \begin{tabular}{@{}lll@{}}
    \toprule   
    Parameters & \xbIII{} & \xbI{} \\   
    \midrule                    
      RA & $6^{\rm h}13^{\rm m}30\fs54$ & $16^{\rm h}12^{\rm m}01\fs98$
      \\     
      DEC & $16\degr04\arcmin42\farcs32$  & $-46\degr46\ \arcmin21\farcs88$   
      \\
      Pos.\ error &  0\farcs98 
      
      &  0\farcs94
      \\
      OBSID  & 0883460301  &  0883460101
      \\
      $T_{\rm start}$, MJD & 59471.18 &  59635.10
      \\
      $T_{\rm stop}$, MJD & 59471.50 &  59635.41
      \\
      Exposure, ks & 27.219 &  26.228
      \\
    \botrule                                 
    \end{tabular}
    \label{xmm_tabl}
\end{table}

\subsection{X-ray positions and optical counterparts}\label{x_ray_pos}
The unambiguous identification of potential optical counterparts in crowded Galactic plane regions requires knowledge of accurate X-ray positions. The accuracy of positions obtained in the eROSITA survey is limited by the counting statistics typical for all-sky surveys. We therefore have used deeper pointed XMM-\textit{Newton} observations to significantly improve the positional accuracy. 
As a first step we determined the positions of all X-ray sources in the field of view using the \texttt{emldetect} task which jointly models data from all three XMM-\textit{Newton} cameras and fully uses the available counting statistics. In order to assess and reduce all remaining systematic uncertainties associated with pointing accuracy, we cross-correlated positions of optical sources detected by XMM-\textit{Newton} optical monitor~(OM) with \textit{Gaia}~DR3 catalogue~\citep{Gaia_miss, Gaiadr3}\footnote{\url{https://cdsarc.cds.unistra.fr/viz-bin/cat/I/355}}. 
The positions of optical sources were determined  using the \texttt{omichain} task. We find 84 and 455 (for \xbIII{} and \xbI{} respectively) objects detected by the OM to have unique \textit{Gaia} counterpart and use those to compute the final astrometrically-corrected positions for both optical and X-ray sources detected by XMM-\textit{Newton} using the \texttt{eposcorr} task. As a result we find
$\alpha_{\rm J2000.0} = 6^{\rm h}13^{\rm m}30\fs54$, $\delta_{\rm J2000.0} = 16\degr04\arcmin42\farcs32$ and $\alpha_{\rm J2000.0}
= 16^{\rm h}12^{\rm m}01\fs98$, $\delta_{\rm J2000.0} = -46\degr46\arcmin21\farcs88$, with an uncertainty of about 0\farcs98, 0\farcs94 at 1-$\sigma$ 
confidence level for \xbIII{} and \xbI{} respectively~(see Table~\ref{xmm_tabl}).

\begin{figure}[t]
\centering
\begin{minipage}{.5\textwidth}
  \centering
  \includegraphics[width=1.15\linewidth]{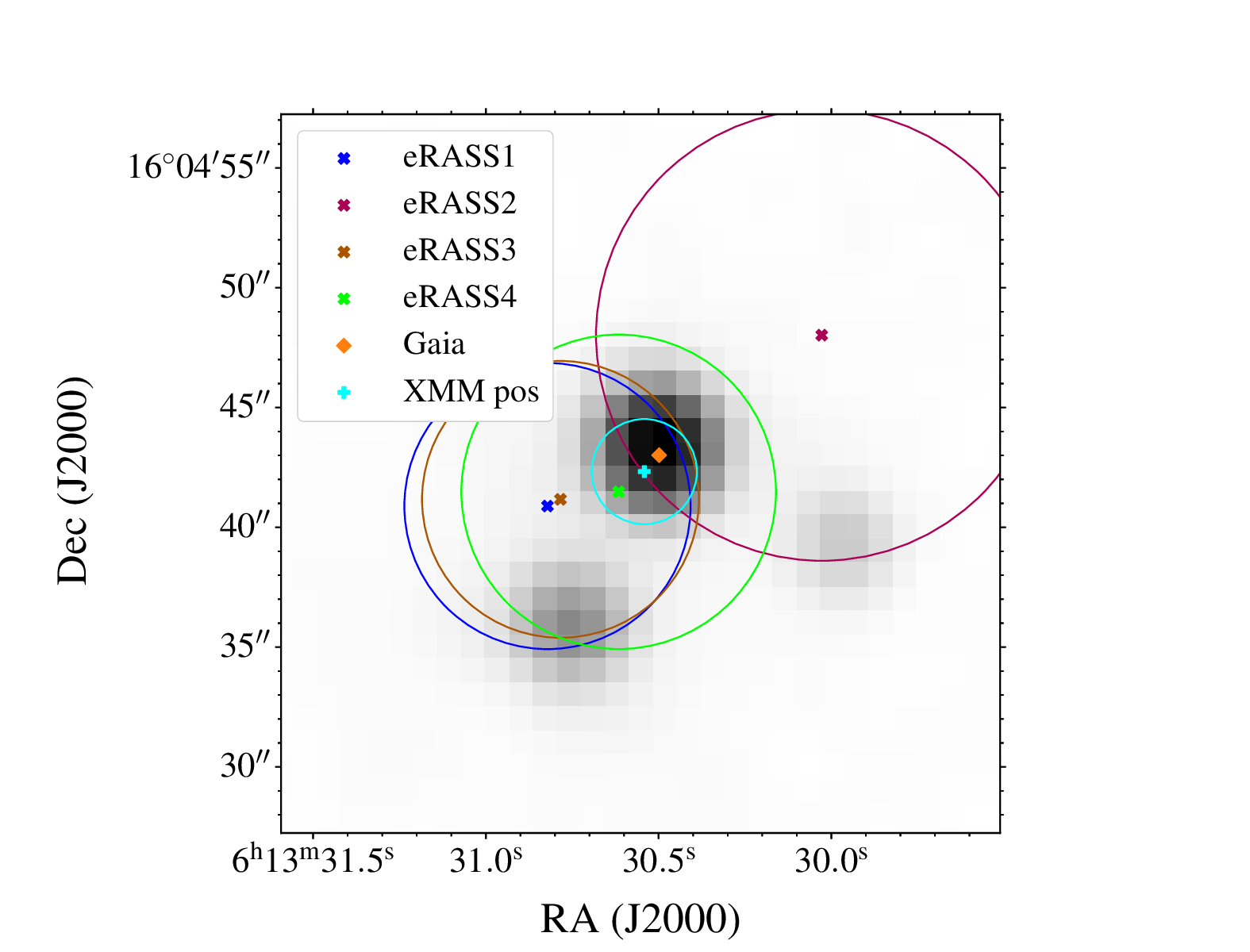}
\end{minipage}%
\begin{minipage}{.5\textwidth}
  \centering
  \includegraphics[width=1.19\linewidth]{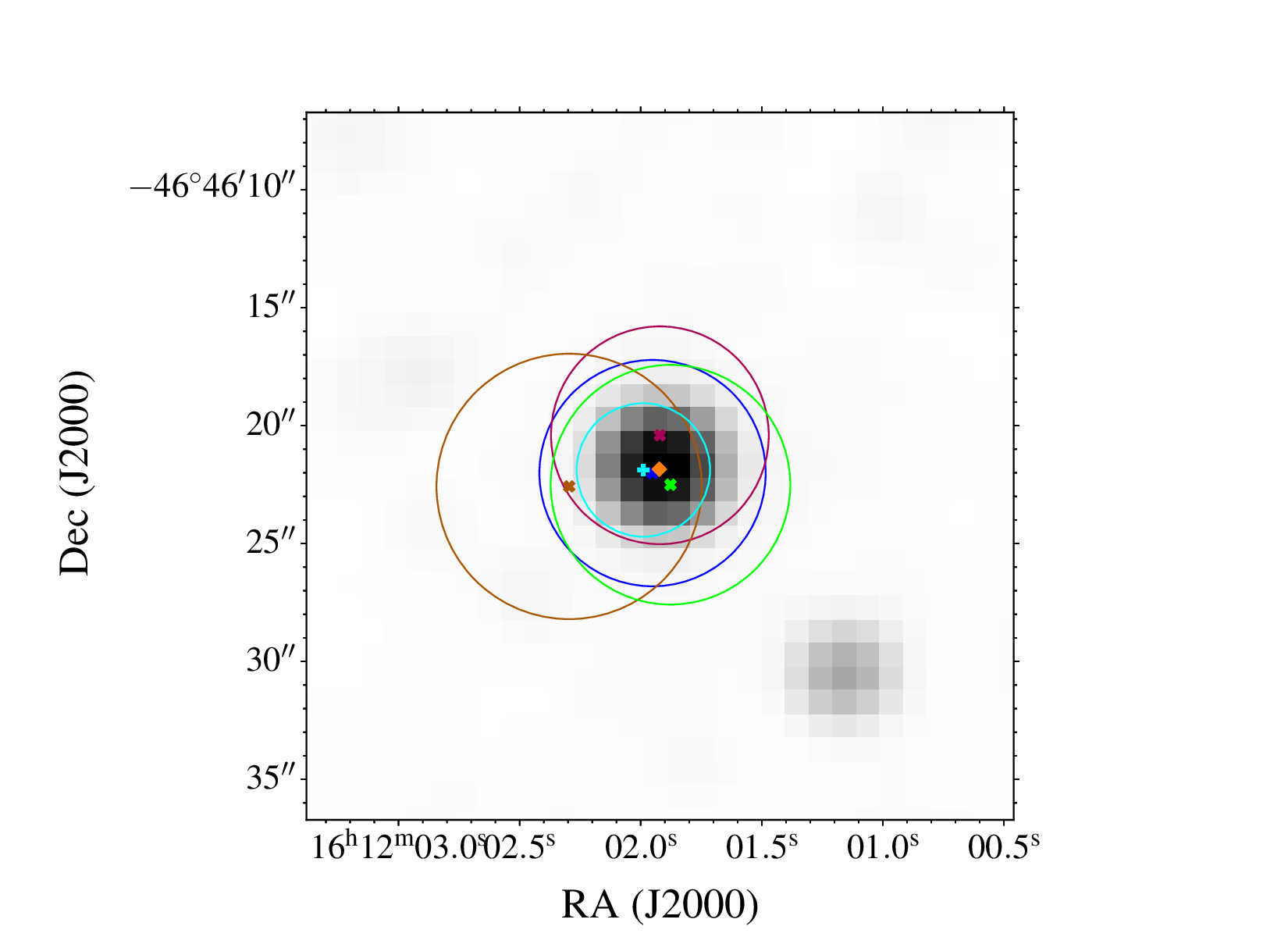}
\end{minipage}
\caption{NIR (2MASS-J) field view around \xbIII{} (left image) and \xbI{} (right image) candidates with eRASS, \textit{Gaia}~DR3 and corrected XMM-\textit{Newton}'s positions. Colour circles show corresponding $3\sigma$ positional errors.}
\label{chart_all}
\end{figure}

Once accurate X-ray positions are determined, one can search for plausible counterparts using optical, near-infrared (NIR) and infrared (IR) catalogues. The NIR 2MASS$-$J~\citep{2MASS} finding charts (obtained through \texttt{astroquery.skyview} package\footnote{\url{https://astroquery.readthedocs.io/en/latest/skyview/skyview.html}}) for \xbIII{} and \xbI{} are presented in the Fig.~\ref{chart_all}~(right and left panel respectively).
Figures show both the astrometrically-corrected XMM-\textit{Newton} positions and the original positions from the eROSITA analysis, which appear to be in good agreement with each other, although the accuracy of the XMM-\textit{Newton} position is of course better. 
The final corrected X-ray positions of both objects appear to have an unique optical/NIR counterparts, and coincide with \textit{Gaia} sources Gaia\,DR3\,3345882675816728960 and Gaia$\,$DR3$\,$5990098842316868352 for \xbIII{} and \xbI{} respectively. These are, therefore, the likely optical counterparts we discuss in detail below.

\subsubsection*{Optical counterpart of \xbIII{}}
The \textit{Gaia} counterpart (Gaia DR3 ID: 3345882675816728960) at an angular separation of 0\farcs924 with the corrected XMM-\textit{Newton} coordinates of \xbIII{}, allows us to estimate the distance to the source at $657^{+58}_{-41}$\,pc~\citep[][hereafter BJ21]{Gaia_dis}\footnote{\url{https://cdsarc.cds.unistra.fr/viz-bin/cat/I/352}}. Further, the star is associated with the known variable star ATO$\,$J093.3770$+$16.0786, which has an average V band magnitude of 15.46\,mag\footnote{\url{https://asas-sn.osu.edu/variables/}} according to the All-Sky Automated Survey for Supernovae~\citep[ASAS-SN,][]{ASAS_SN, ASAS_V} catalogue.
The source is also monitored by the Zwicky Transient Facility~\citep[ZTF, we use DR8,][]{ZTF}. Inspection of the ZTF data using the SNAD ZTF viewer\footnote{\url{https://ztf.snad.space/}}~\citep{ZTF_per, ZTF_per2} reveals periodic variability with period of $7.189$ days detected with signal-to-noise ratio~$S/N=22.891$. 
The Asteroid Terrestrial-impact Last Alert System (ATLAS) catalogue~\citep{ATLAS, ATLAS_V}\footnote{\url{https://cdsarc.cds.unistra.fr/viz-bin/cat/J/AJ/156/241}}, on the other hand, also indicates the variable nature of the star but \boldt{provides several different values for the possible period, namely: 1.157, 2.315, and 7.211\,days. The first two periods seem to be multiplicities of one another caused by aliasing. Furthermore, the ATLAS variable stars catalogue puts ATO$\,$J093.3770$+$16.0786 into the NSINE (noisy sine wave) category, which indicates the presence of sinusoidal variables with strong residual noise or with evidence of additional variability not captured in the fit.}

\subsubsection*{Optical counterpart of \xbI{}}

For the second object, we identify the only plausible counterpart 0\farcs679 away from the corrected XMM-\textit{Newton} position of \xbI{}, Gaia$\,$DR3$\,$5990098842316868352. The distance to this source is estimated to be $392^{+2}_{-2}$\,pc~(see \textit{Gaia}~DR3). \boldt{The substantially more precise distance estimate of J161201 compared to J161331 is due to its closer proximity to us, resulting in a larger parallax (2.53\,mas vs.\ 1.52\,mas) and consequently smaller uncertainty. Moreover, \xbI{} is significantly brighter in the optics than \xbIII{} (12.32\,mag vs.\ 15.31\,mag in the $G$-band), so \textit{Gaia}~DR3 does not provide a distance estimate with their BP/RP spectra Aeneas algorithm for \xbIII{}'s counterpart. Consequently, for \xbIII{} we used values from BJ21, despite higher uncertainties arising from their method, which incorporates parallax, photometric measurements, and fitting the distribution of stars in the Milky Way.} 

The optical counterpart of \xbI{} is also a known variable star in the ASAS-SN catalogue with an average V band magnitude of $12.42$ and a tentatively identified period of about 177.8 days. \boldt{Additionally, we looked for Transiting Exoplanet Survey Satellite~\citep[TESS,][]{2015JATIS...1a4003R} catalogue data, and found that the star is flagged as a binary candidate with the suggested periodicity of 7.0753\,days~\citep{2023MNRAS.522...29G}\footnote{\url{https://cdsarc.cds.unistra.fr/viz-bin/cat/J/MNRAS/522/29 }}.} The object is also present in the AllWISE catalogue, where it is identified as J161201.92$-$464621.8. The catalogue indicates variability in three out of four of WISE's mid-infrared bands.

\boldt{A summary of the optical and distance information regarding the suggested counterparts to \xbI{} and \xbIII{} from \textit{Gaia}~DR3 and BJ21, along with NIR JHK information from 2MASS is presented in Table~\ref{opt_tabl}.}

\begin{table}[t]
\renewcommand{\arraystretch}{1.4} 
\renewcommand{\tabcolsep}{2mm}   
    \centering
    \caption{Summary of optical/IR and distance information about the suggested counterparts.}        
    \begin{tabular}{@{}llll@{}}
    \toprule   
    Parameters & Description & \xbIII{} & \xbI{} \\   
    \midrule                    
      G, mag & Mean apparent G-band magnitude & 15.3089$^{+0.0051}_{-0.0051}$  & 12.3203$^{+0.0067}_{-0.0067}$  
      \\     
      BP-RP, mag & BP-RP colour & 2.3561$^{+0.0178}_{-0.0178}$  & 0.9392$^{+0.0395}_{-0.0395}$
      \\
      A$_{\rm V}$, mag & Extinction in V band  & --- & 0.4429$^{+0.0098}_{-0.0096}$	 
      \\
      A$_{\rm G}$, mag & Extinction in G band  & --- & 0.3717$^{+0.0084}_{-0.0081}$
      \\
      J, mag & Apparent J-band magnitude & 11.555$^{+0.022}_{-0.022}$  & 10.825$^{+0.023}_{-0.023}$  
      \\   
      H, mag & Apparent H-band magnitude & 10.598$^{+0.021}_{-0.021}$  & 10.539$^{+0.023}_{-0.023}$  
      \\
      K, mag & Apparent K-band magnitude & 10.169$^{+0.018}_{-0.018}$  & 10.330$^{+0.019}_{-0.019}$  
      \\   
      D, pc & Distance & $656.9^{+57.5}_{-40.5}$ & $391.7^{+2.2}_{-2.1}$
      \\
    \botrule                                 
    \end{tabular}
    \label{opt_tabl}
\end{table}

\section{Data analysis results}\label{data_anal}
\subsection{X-ray spectral analysis}\label{spectra}

Considering the low statistics in eROSITA spectra, we combined the data from all four individual surveys. For XMM-\textit{Newton}, the spectra from each of the EPIC cameras (MOS1, MOS2, PN) were extracted separately and analysed together with the combined eROSITA spectrum using the same model. A cross-normalisation constant was introduced for each of the spectra to account for differences in the absolute flux calibration of individual instruments and possible flux differences arising from asynchronous XMM-\textit{Newton} and eROSITA observations. All spectra were grouped to contain at least one count per energy bin and were modelled using XSPEC v12.11.1\footnote{\url{https://heasarc.gsfc.nasa.gov/xanadu/xspec/}}~\citep{Xspec}.

\begin{figure}[t]
\centering
\begin{minipage}{.5\textwidth}
  \centering
  \includegraphics[width=0.97\linewidth]{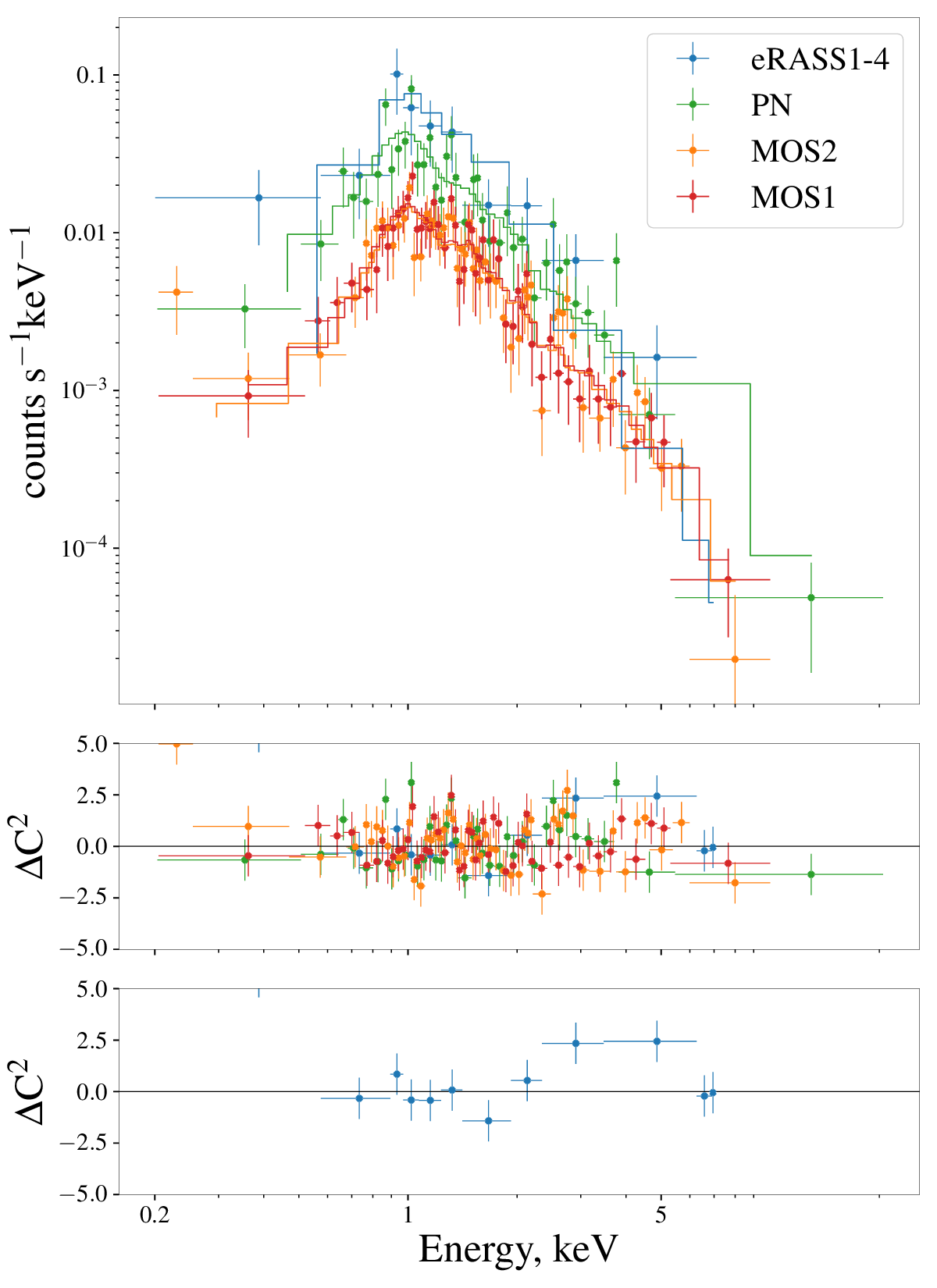}
\end{minipage}%
\begin{minipage}{.5\textwidth}
  \centering
  \includegraphics[width=0.97\linewidth]{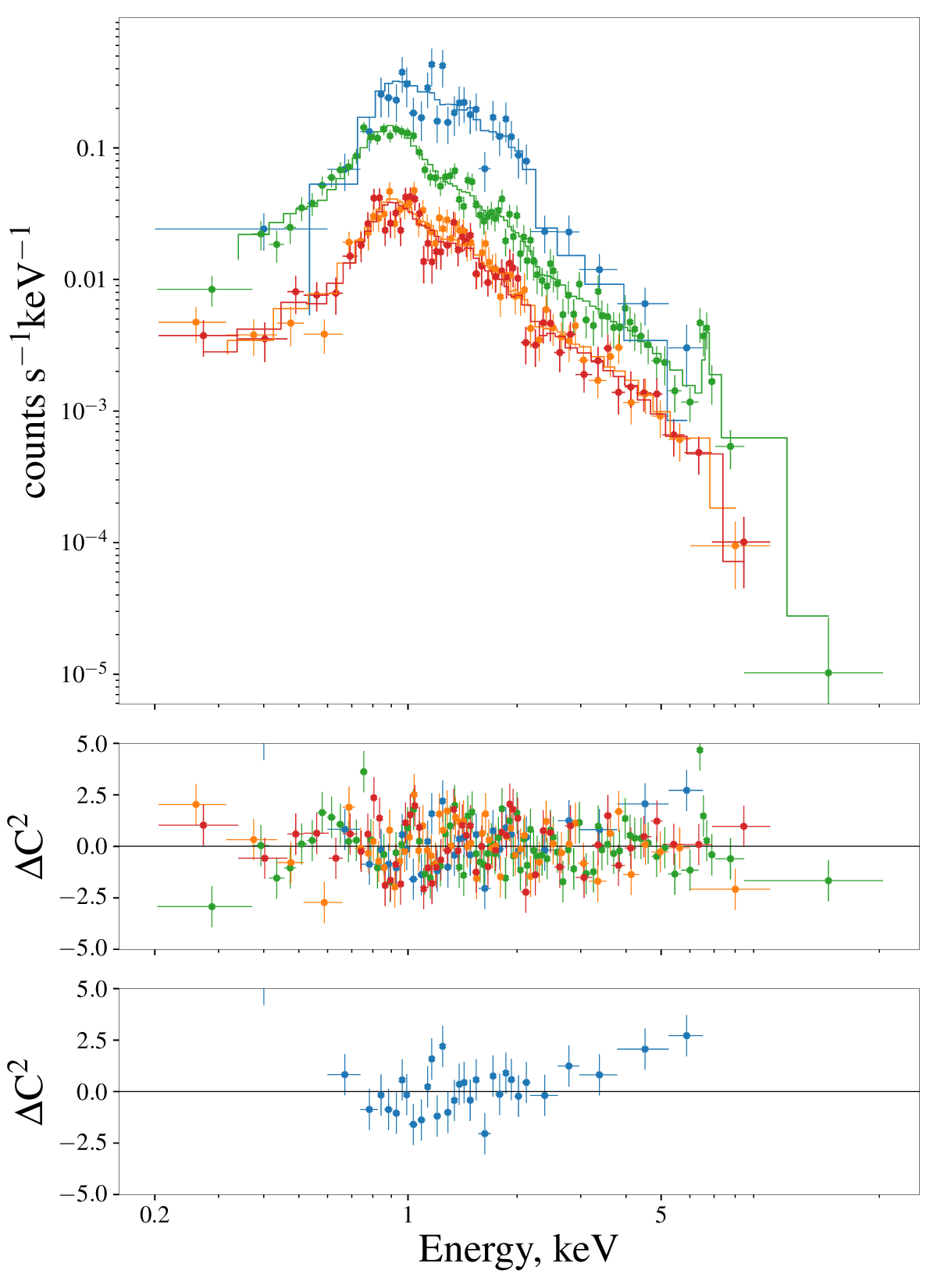}
\end{minipage}
 \caption{XMM-\textit{Newton} and eROSITA spectra of \xbIII{} (left image) and \xbI{} (right image) fitted with double apec~(\texttt{const*tbabs*(apec+apec)}) model. \textit{Top panels:} spectra themselves and the fit; \textit{Middle panels:} all corresponding residuals; \textit{Bottom panels:} eROSITA residuals only.}
\label{Spec_all}
\end{figure}

To model the spectra we used the phenomenological models commonly used to describe the spectra of XRBs and active stars. First, we attempted to model the spectra using absorbed cutoff power law (\texttt{constant*tbabs*cutoffpl}) and absorbed black body models. To take into account interstellar absorption we use the \texttt{TBabs} component~\citep{Wilms2000}. We found that both models fail to provide a statistically acceptable approximation for the spectra of either source. 
On the other hand, we found that an acceptable fit can be obtained using a two-temperature optically thin thermal plasma emission model~\citep{1977ApJS...35..419R,2001ApJ...556L..91S}, i.e. \texttt{constant*tbabs*(apec+apec)}
commonly used to approximate the spectra of hot~\cite{Naze14} and cool~\cite{2005A&A...435.1073R, Stelzer_er} stars, as well as some accreting white dwarfs (WDs) a.k.a. cataclysmic variables (CVs)~\citep{2020A&A...639A..17W,2004MNRAS.350.1373R}. The best-fit results and residuals are presented in Fig.~\ref{Spec_all} and Table~\ref{tab:specparams}.

\begin{table}[t]
\renewcommand{\arraystretch}{1.4} 
\renewcommand{\tabcolsep}{2mm}   
\centering
\caption{\xbIII{} and \xbI{} parameters of double apec spectral model \texttt{const*tbabs*(apec+apec)} used to fit the XMM-\textit{Newton} and eROSITA spectra. Abundances are fixed to solar.}
\begin{tabular}{@{}llll@{}}
\toprule  
Components & Parameters & \xbIII{} & \xbI{}
\\
\midrule   
\texttt{constant} & $\mathrm{C_{MOS1}}$ & 1.0 & 1.0 \\
& $\mathrm{C_{MOS2}}$ & $0.93^{+0.13}_{-0.11}$ & $1.02^{+0.09}_{-0.08}$     \\
& $\mathrm{C_{PN}}$ & $0.80^{+0.12}_{-0.11}$ &  $0.99^{+0.07}_{-0.07}$  \\
& $\mathrm{C_{eROSITA}}$ & $2.07^{+1.34}_{-0.93}$ &  $6.62^{+1.32}_{-1.10}$  \\
\texttt{tbabs} & $N_\mathrm{H_1, XMM}, 10^{22}\,\mathrm{cm^{-2}}$ & $0.33^{+0.10}_{-0.07}$ & $0.22^{+0.03}_{-0.03}$
\\
& $N_\mathrm{H_1, eROSITA}, 10^{22}\,\mathrm{cm^{-2}}$ & $0.55^{+0.37}_{-0.34}$ & $0.71^{+0.13}_{-0.11}$
\\
$\mathtt{apec_1}$ & $kT_1$, keV & $0.87^{+0.11}_{-0.09}$ & $0.94^{+0.04}_{-0.04}$ \\
&  $\mathrm{norm_1}$, $10^{-5}$\,cm$^{-5}$ & $3.21^{+1.33}_{-1.00}$ & $6.69^{+0.94}_{-0.82}$ 
\\
$\mathtt{apec_2}$ & $kT_2$, keV & $3.46^{+0.87}_{-0.73}$ & $5.84^{+1.48}_{-0.81}$ 
\\
&  $\mathrm{norm_2}$, $10^{-5}$\,cm$^{-5}$ & $12.86^{+1.97}_{-1.86}$ & $22.91^{+1.77}_{-1.70}$ 
\\
\botrule    
\end{tabular}
\label{tab:specparams}
\end{table}

\begin{figure}[t]
   \centering
   \includegraphics[width=0.65\textwidth]{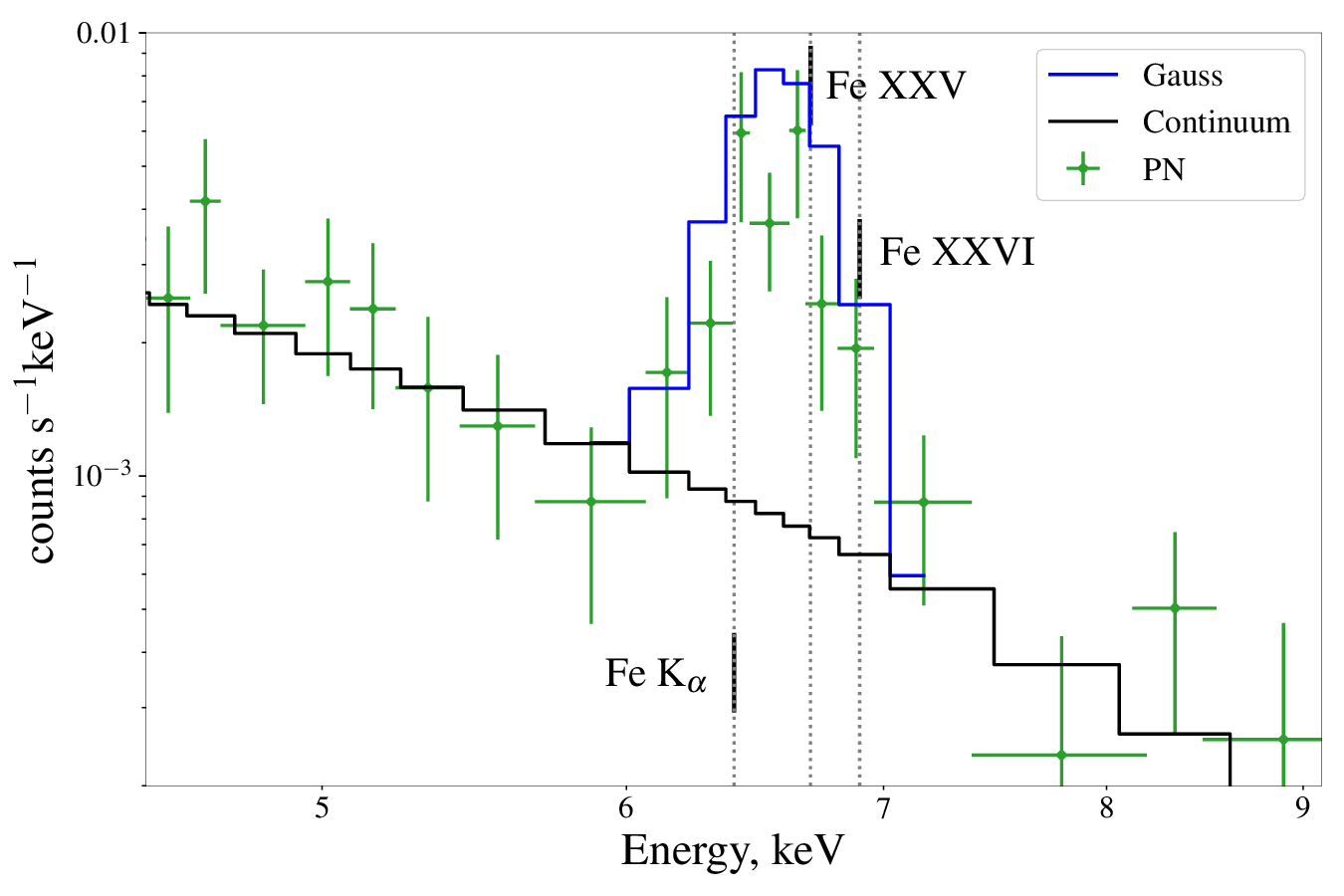}
  \caption{Zoom-in region of XMM-\textit{Newton} spectra for \xbI{} (PN camera only) with shown continuum level and Gaussian fit. Vertical lines show the positions of corresponding main iron emission lines (6.4, 6.7 and 6.9 keV respectively).}
    \label{Spec_Fe}
   \end{figure}

\boldt{The X-ray spectrum of \xbI{}~(see Fig.~\ref{Spec_all} and Table~\ref{tab:specparams}) appears to be harder than that of \xbIII{}. More importantly, a closer look at PN data for \xbI{} in the Fig.~\ref{Spec_all} (right panel), indicates the presence of an iron emission line around 6-7\,keV, which is missed in MOS1-2 data likely due to lower statistics. A zoomed view of the region around the emission line is shown in Fig.~\ref{Spec_Fe}. The results of a Gaussian fit to the line are presented in the Table~\ref{tab:line}.}
    
\boldt{In general, for X-ray sources, iron line emission in the 6--7\, keV energy range constitutes a set of distinct lines, consisting primarily of three components: a fluorescent Fe~K$_{\alpha}$ line at 6.4 keV, and two Ly$_{\alpha}$ lines from Fe XXV (He-like iron) near 6.7\,keV and Fe XXVI (H-like iron) around 6.9\,keV. These lines are essential markers of hard X-ray emission acting in sources such as XRBs~\citep{1978ApJ...223..268B,2000ApJS..131..571A,2015A&A...576A.108G,2010ApJ...715..947T}, CVs~\citep{2005ApJ...626..396P,2012MmSAI..83..585B,2006ApJ...642.1042R,2014MNRAS.437..857E,2020A&A...639A..17W}, AGNs~\citep{1978ApJ...220..790M,1990ApJ...360L..35P,1990ApJ...361..440M,1990Natur.344..132P} and some RS Canum Venaticorum (RS~CVn) variables~\citep{1981MNRAS.196P..73A,2021ApJ...910...25S,2024ApJ...969L..12I}. As fluorescent lines, these lines are a measure for the X-ray flux above the K-edge of Fe. In turn, Ly$_{\alpha}$ lines come from photoionization or collisional excitation in hot plasma with a temperature of $10^{7\ldots8}$ K.}

\boldt{However, poor statistics do not allow us to resolve each iron line separately. Nevertheless, the fact that we do see iron line emission around 6--7\,keV indicates the presence of hard X-ray radiation, which is not typical for regular active stars.}

    \begin{table}[ht!]   
    \renewcommand{\arraystretch}{1.4} 
    \renewcommand{\tabcolsep}{2mm}   
    \centering
    \caption{Gaussian fit parameters for the iron emission line of \xbI{} spectra.}
    \begin{tabular}{@{}ll@{}}
    \toprule  
    Parameters & Values
    \\
    \midrule   
    Line energy, keV & 6.55 $\pm$ 0.03 
    \\
    Line normalisation, $10^{-5}$\,photons\,cm$^{-2}$\,s$^{-1}$ & 3.08 $\pm$ 0.15
    \\
    Full width at half maximum (FWHM), eV  & 435 $\pm$ 23 
    \\
    \botrule  
    \end{tabular}
    \label{tab:line}
    \end{table}

\subsection{X-ray timing analysis}\label{timing}
\begin{figure}[t]
\centering
\includegraphics[width=1\textwidth]{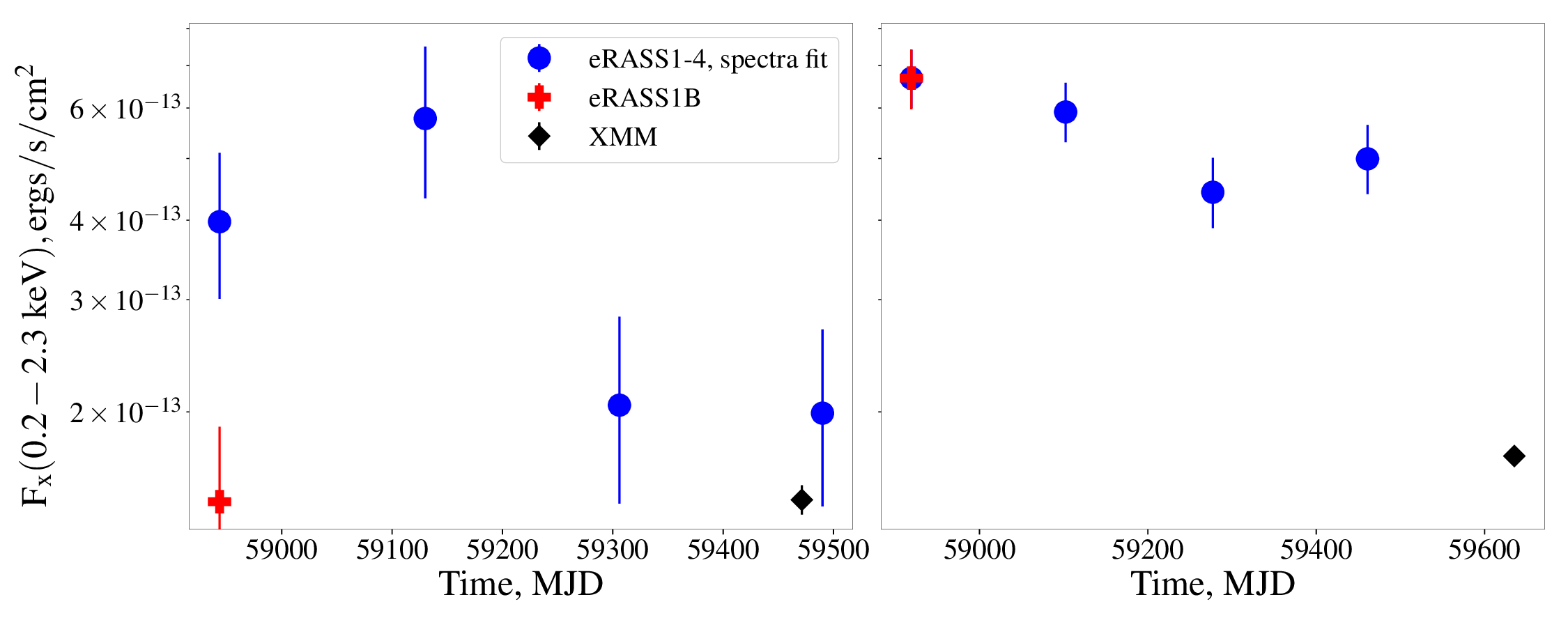}
\caption{Intrinsic X-ray fluxes of \xbIII{}~(left panel) and \xbI{}~(right panel) based on XMM-\textit{Newton} and eROSITA data~(both from spectra fitting and from eRASS's data).}
\label{Xray_lc}
\end{figure}

To assess the long-term X-ray variability of the source we estimated the source flux by modelling source spectra extracted for each of the individual observations and approximated using the same model where all parameters except the normalisation were fixed to their best fit values. To estimate observed flux and its uncertainty the additional \texttt{cflux}~(as \texttt{cflux*tbabs*(apec+apec)}) model component (with an energy range 0.2--2.3 keV for eROSITA and XMM-\textit{Newton}) was introduced. The energy range was chosen to match that of eRASS1 catalogue\footnote{\url{https://cdsarc.cds.unistra.fr/viz-bin/cat/J/A+A/682/A34}}.
The flux value was considered to be independent for each observation. The resulting long-term X-ray light curve of the source is presented in Fig.~\ref{Xray_lc}.

We have also carried out a search for possible coherent pulsations for both objects using the XMM-\textit{Newton} data. The periodicity search was carried out using the observed event arrival times (corrected to Solar system barycentre) and $Z^{2}$ statistics~\citep{Jager89} as implemented in the \texttt{Stingray}\footnote{\url{https://docs.stingray.science/}} package~\citep{Stingray}. We used the energy range (0.5-10\,keV) and separately considered two ranges of possible periods for the full data set (all three cameras combined and PN camera only) to exploit finer time resolution of the latter instrument. We did not search for pulsations in the eROSITA data due to the low number of counts there. The range of periods searched is, therefore, defined by the duration of the observations and Nyquist frequency $\nu_{\rm Nyquist}$ corresponding to time resolution $ t_{\rm res}$ of a given detector (in particular $\nu_{\rm Nyquist}=\mathrm{0.5 \times t_{\rm res}}$). The resolution of the periodograms was chosen to oversample the latter by a factor of 8 to avoid missing any peaks. In each case, periodograms were averaged over segments with a length exceeding the longest period searched by a factor of 200 to reduce noise and further improve sensitivity. The ranges of periods searched and number of photons used in each search are listed in Table~\ref{tab:period}.

\begin{table}[ht!]
\renewcommand{\arraystretch}{1.4} 
\renewcommand{\tabcolsep}{2mm}   
\centering
\caption{Parameters of pulsation search for \xbIII{} and \xbI{}.}
\begin{tabular}{@{}llll@{}}
\toprule 
Parameters & Camera & \xbIII{} & \xbI{} \\
\midrule  
Counts & all EPIC & 1828 & 5041  \\
& PN only & 379 & 2945  \\
Periods range, s  & all EPIC & 5.2-27219 & 5.2-26228 \\
& PN only & 0.1468-22055 & 0.1468-21370 \\
$f_{\rm puls}$ limit, \%  & all EPIC & 23.37 & 19.01 \\
& PN only  &  36.23 & 25.66 \\
\botrule  
\end{tabular}
\label{tab:period}
\end{table}

Unfortunately, no evidence of significant coherent X-ray periodicity was found for either candidate. Considering the relatively low flux of both sources, this could be, however, simply a result of insufficient counting statistics, so we also estimated the upper limits on the amplitude of a possible coherent signal which could be missed by our search. In particular, we used the methods suggested by \cite{Brazier94} as implemented and described in detail in \cite{Doroshenko15puls} and by \cite{Vaughan94} as implemented in \texttt{amplitude\_upper\_limit} and
\texttt{pf\_from\_a\_stingray} functions in \texttt{Stingray} package.

The resulting upper limits for 3-$\sigma$ confidence level are also listed in Table~\ref{tab:period}. The lack of a clearly detected pulse signal is not surprising, given the available statistics for this observation, so no conclusions regarding the nature of either source shall be drawn based on the non-detection of the pulsations.

\subsection{Optical spectroscopy}\label{opt_spectra}

\boldt{To complement X-ray observations, we attempted to obtain archival spectral data in optical/IR bands. While there is no archival optical/IR spectral information available for \xbI{}, data from \xbIII{} can be found in the Large sky Area Multi-Object fiber Spectroscopic Telescope~\citep[LAMOST,][]{2012RAA....12.1197C} archive. This source has been classified as a late-type star with a spectral type between K7~\citep{2019ApJS..244....8Z}\footnote{\url{https://cdsarc.cds.unistra.fr/viz-bin/cat/J/ApJS/244/8}} and M0~\citep{2021ApJS..253...19Z}\footnote{\url{https://cdsarc.cds.unistra.fr/viz-bin/cat/J/ApJS/253/19}}. In Table~\ref{tab:lam} we also give other derived quantities for the star as obtained from LAMOST.}

\subsubsection*{LAMOST optical spectrum of \xbIII{}}

\begin{figure}[t]
\centering
\includegraphics[width=1\textwidth]{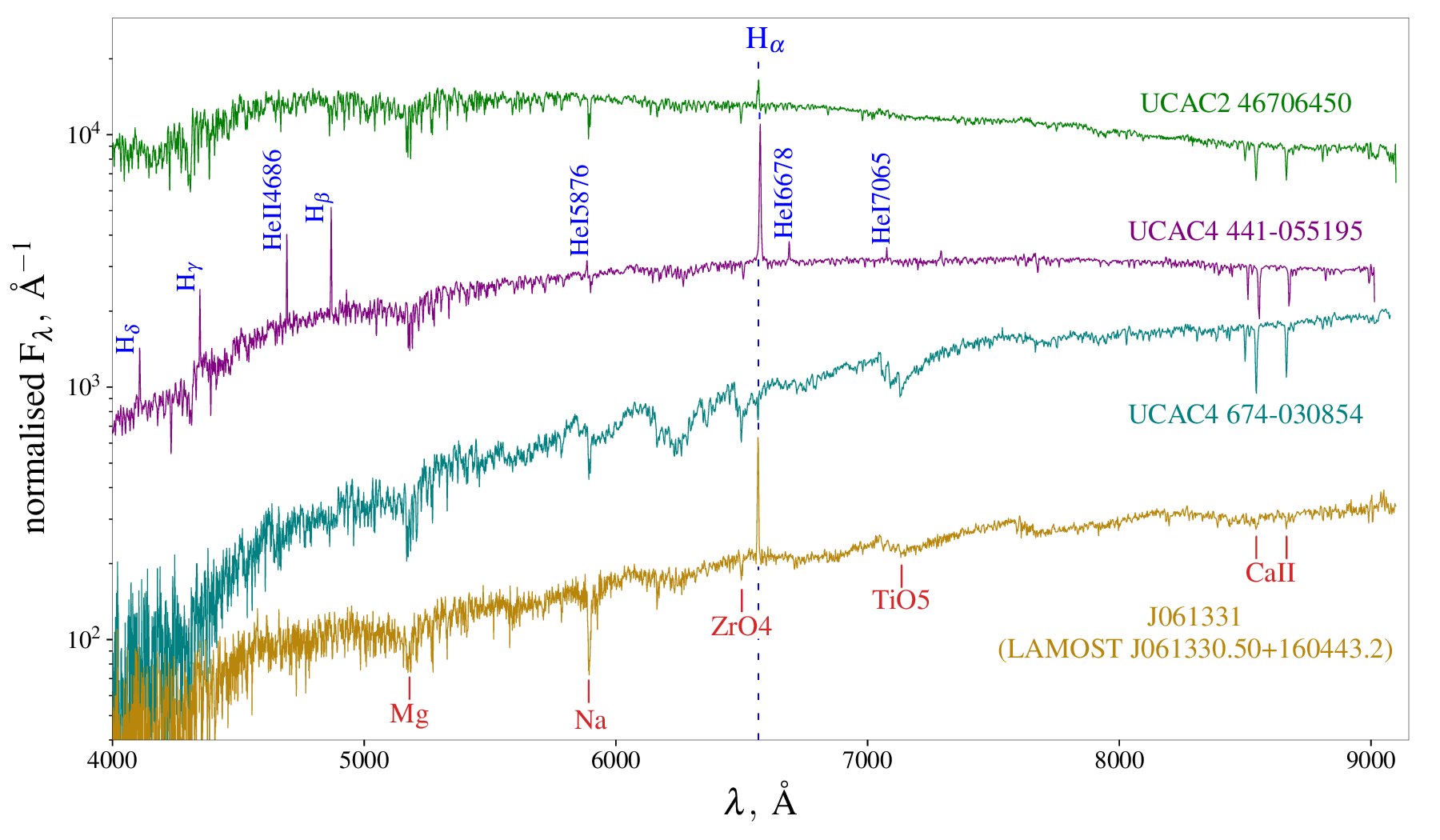}
\caption{\boldt{From the bottom to the top: LAMOST DR9 low-resolution ($ \lambda/\Delta \lambda = 1800$) optical spectra of the counterpart to \xbIII{}, UCAC4\,674$-$030854 (S-type giant), UCAC4\,441$-$055195 (Symbiotic CV with K-type giant) and UCAC2\,46706450 (K-type subgiant with non-accreting WD). Spectra of UCAC4\,674-030854 and UCAC4\,441$-$055195 were rescaled for ease of plotting. Strong emission and absorption lines are marked with blue and red coloured text, respectively.}}
\label{xrb3_optspectra}
\end{figure}

\boldt{The corresponding LAMOST optical spectra\footnote{\url{https://www.lamost.org/dr9/}} are presented in Fig.~\ref{xrb3_optspectra}.  Along with the counterpart of \xbIII{}, we show spectra of several other sources of different origins which appear in LAMOST low-resolution data, namely UCAC4\,674$-$030854~\citep[S-type giant,][]{2022ApJ...931..133C,2023ApJS..267....5C}\footnote{\url{https://cdsarc.cds.unistra.fr/viz-bin/cat/J/ApJ/931/133}}$^{,}$\footnote{\url{https://cdsarc.cds.unistra.fr/viz-bin/cat/J/ApJS/267/5}}, UCAC4\,441$-$055195~\citep[Symbiotic CV with K-type giant,][]{2015RAA....15.1332L}, and UCAC2\,46706450~\citep[K-type subgiant with non-accreting WD,][]{2020A&A...642A.228W}.}

\boldt{As is evident from Fig.~\ref{xrb3_optspectra}, the LAMOST spectrum of \xbIII{}'s optical counterpart shows a very strong $\mathrm{H_{\alpha}}$ emission line (6564.6\,$\mathrm{\r{A}}$). The $\mathrm{H_{\alpha}}$ emission line is an essential indicator of rotation and chromospheric/coronal activity for late-type stars~\citep{1990ApJS...72..191S,2014ApJ...795..161D,2017ApJ...834...85N,2019A&A...628A..41P}, accreting processes for young stellar objects~(YSO)~\citep{1945ApJ...102..168J,2003ApJ...592..282J,2005ApJ...620L..51L}, Be-type~\citep{2013A&A...559A..87Z,2019A&A...622A.173Z} or symbiotic~\citep{1977ApJ...211..866D} XRBs, as well as CVs~\citep{2013MNRAS.432.3186M}. We calculate the corresponding FWHM of $\mathrm{H_{\alpha}}$ line for \xbIII{}'s optical counterpart to be $7.4\pm0.3$\,$\mathrm{\r{A}}$~(also presented in Table~\ref{tab:lam}). However, we do not see other emission signatures, for example, $\mathrm{H_{\beta}}$,  $\mathrm{H_{\gamma}}$, He lines typical for accreting binaries with compact companions~\citep{2013MNRAS.432.3186M,2016MNRAS.460..513T,2023MNRAS.521.4190K,2024A&A...690A.148A}. On the other hand, even though LAMOST detects TiO and other absorption lines (such as Mg, NA, etc.), which come from the star's photosphere, those are relatively weak making it less likely for \xbIII{} to be a giant star and rather be still on subgiant stage~(see the comparison with UCAC4\,674$-$030854 and UCAC2\,46706450 in the Fig.~\ref{xrb3_optspectra}).}

\begin{table}[ht!]
\renewcommand{\arraystretch}{1.4} 
\renewcommand{\tabcolsep}{2mm}   
\centering
\caption{The parameters of \xbIII{}'s optical counterpart based on LAMOST spectrum.}
\begin{tabular}{@{}lllllll@{}}
\toprule 
Parameter & $T_{\rm eff}$, K &  $\log(g)$, cm/s$^2$ & M/H, dex & $\mathrm{FWHM_{H\alpha}}$, $\mathrm{\r{A}}$ &  $[\mathrm{TiO}]_{1}$ & $[\mathrm{TiO}]_{2}$
\\
\midrule 
Value & $3817\pm101$ & $4.41\pm 0.22$ & $-0.91 \pm 0.28$ & $7.4\pm0.3$ &  0.022 &  0.012 
\\
\botrule  
\end{tabular}
\label{tab:lam}
\end{table}

\boldt{In order to find the spectral type of \xbIII{}'s optical counterpart, we use methods presented in~\citep{1987AJ.....93..938K}. Based on a large dataset of late-type stars authors found that the depth of TiO
bands could be utilised as an indicator for star's spectral type, namely~\citep[see equations 5 and 6 in][]{1987AJ.....93..938K}:
\begin{equation}\label{equs1}
\mathrm{ST}_{1} = 1.75 + 9.31\,[\mathrm{TiO}]_{1}\,,
\end{equation}
\begin{equation}\label{equs2}
  \mathrm{ST}_{2} = 1.83 + 10.37\,[\mathrm{TiO}]_{2} - 3.28\,[\mathrm{TiO}]_{2}^2\,,
\end{equation} 
where the $\mathrm{ST}_{1(2)}$ is a spectral type, while [TiO]$_{1}$ and [TiO]$_{2}$ are corresponding TiO indices~\citep[see equations 1 and 2 in][]{1987AJ.....93..938K}. $\mathrm{ST}_{1(2)}$ lies around $0$ for M0 stars, $-6$ for K0 stars, and $+6$ for M6 stars. Following equations 1 and 2 from  \cite{1987AJ.....93..938K}, we calculate TiO indices based on the LAMOST spectra for \xbIII{}'s optical counterpart~(see Table~\ref{tab:lam}). Therefore, we determine ST$_1$ and ST$_2$ for \xbIII{}'s optical counterpart to be both around $1.9$, resulting in the spectral type of the star close to M2. This is a bit later than what LAMOST archival data suggest (but does not contradict it).}

\subsubsection*{Follow-up spectroscopy of \xbI{} with SALT}\label{spec_xb1}

\begin{figure}[t]
    \centering
    \includegraphics[width=\textwidth]{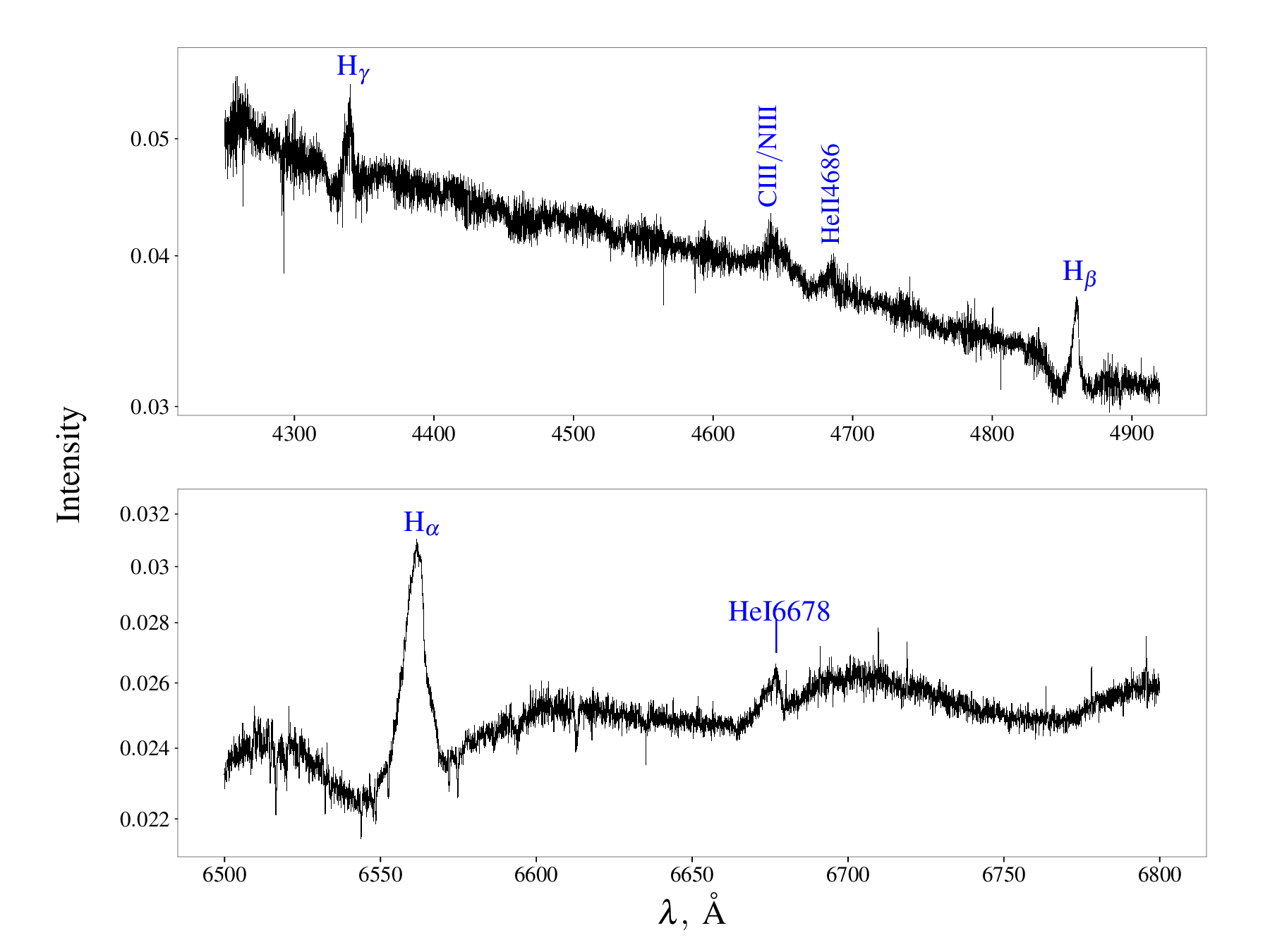}
    \caption{Blue (top panel) and red (bottom panel) arms of \xbI{}'s SALT HRS spectrum. SALT HRS provides data in arbitrary, uncalibrated intensity units. Wavelength regions are selected to showcase identified emission lines.}
  \label{xrb1_optspectra}
\end{figure}

\boldt{Since no archival spectral data were available for the optical counterpart of \xbI{}, the source was selected for the follow-up with the 10\,m class Southern African Large Telescope~\citep[SALT,][]{2006SPIE.6267E..0ZB}. A 1600\,s observation was taken with SALT's High Resolution Spectrograph~\citep[HRS,][]{2008SPIE.7014E..0KB,2010SPIE.7735E..4FB,2012SPIE.8446E..0AB,2014SPIE.9147E..6TC} on February 5.1026, 2025 in low-resolution (LR; R $\sim$ 15000) mode. HRS is a dual-beam, fibre-fed échelle spectrograph that spans a wavelength range of 3800 to 8900\,$\mathrm{\r{A}}$. The initial processing of the HRS spectra was carried out with the PySALT software package~\citep{2010SPIE.7737E..25C}\footnote{\url{http://pysalt.salt.ac.za/}} which performs tasks such as overscan removal, bias correction, and gain calibration. Spectral extraction was conducted using the HRS pipeline, which utilizes MIDAS routines~\citep{2016MNRAS.459.3068K}. The resulting spectrum is shown in Fig.~\ref{xrb1_optspectra}.}

\begin{table}
\renewcommand{\arraystretch}{1.4} 
\renewcommand{\tabcolsep}{2mm}   
\centering
\caption{Spectral line measurements for \xbI{} SALT spectrum. EW and FWHM values are shown for emission and absorption (in case of H-Balmer) line profiles.}
\begin{tabular}{@{}lllllll@{}}
\toprule 
Line / Parameter  & EW, \AA        & FWHM, km/s         & Abs. EW, \AA     & Abs. FWHM, km/s       \\
\midrule
H-alpha           & $-2.81 \pm 0.03$ & $380 \pm 30$   & $4.21 \pm 0.16$   & $2150 \pm 120$    \\
H-beta            & $-1.08 \pm 0.02$ & $370 \pm 20$   & $2.36 \pm 0.07$   & $2350 \pm 150$    \\
H-gamma           & $-1.06 \pm 0.06$ & $450 \pm 70$  & $1.66 \pm 0.11$   & $1780 \pm 270$   \\
He\,I\,6678       & $-0.22 \pm 0.01$ & $330 \pm 20$  & N/A              & N/A                    \\
C\,III/N\,III     & $-1.48 \pm 0.08$ & $1500 \pm 70$ & N/A              & N/A                    \\
He\,II\,4686      & $-0.19 \pm 0.01$ & $360 \pm 30$  & N/A              & N/A                    \\
\botrule
\end{tabular}
\label{tab:SALT_em}
\end{table}

\boldt{The SALT spectrum obtained for \xbI{} displays a very blue continuum, which continues to rise toward the short-wavelength cut-off near 3800\,\AA. This suggests a strong ultraviolet component and hints at high temperatures in the emitting region. The spectrum is rich in emission features, prominently showing Balmer series lines such as H$_\alpha$, H$_\beta$, and H$_\gamma$ (see Fig.~\ref{xrb1_optspectra}). In addition to H-Balmer emission lines the source displays He-emission lines (neutral He\,I\,6678 and ionised He\,II\,4686) as well as fluorescent C\,III/N\,III Bowen blend~\citep[fluorescent for N\,III, see][]{1935ApJ....81....1B,1992ApJ...389L..63F} around 4640\,\AA, which is often observed in accreting X-ray binaries~\citep{1975ApJ...198..641M}, such as LMXBs~\citep{2008AIPC.1010..148C} and CVs~\citep{1991ApJ...373..633S,1999MNRAS.305..437H}. EWs and FWHMs of emission and absorption line profiles are shown in Table~\ref{tab:SALT_em}. The nature of the optical spectrum robustly points toward a CV classification for \xbI{}. The resulting spectrum shows no evidence of the donor star. The humps and the overall rise in intensity towards higher wavelengths seen in red arm of the SALT HRS data~(Fig.~\ref{xrb1_optspectra}, bottom panel) are due to the échelle design of the spectrograph\footnote{\url{https://astronomers.salt.ac.za/wp-content/uploads/sites/71/2014/08/3200-AE-0017_SALT_HRS_Instrument_Description_v4.0.pdf}} and the lack of flux calibration --- not the photosphere of the secondary star.}

\subsection{Optical variability}\label{optvar}

\begin{figure}[t]
\centering
\begin{minipage}{.5\textwidth}
  \centering
  \includegraphics[width=1\linewidth]{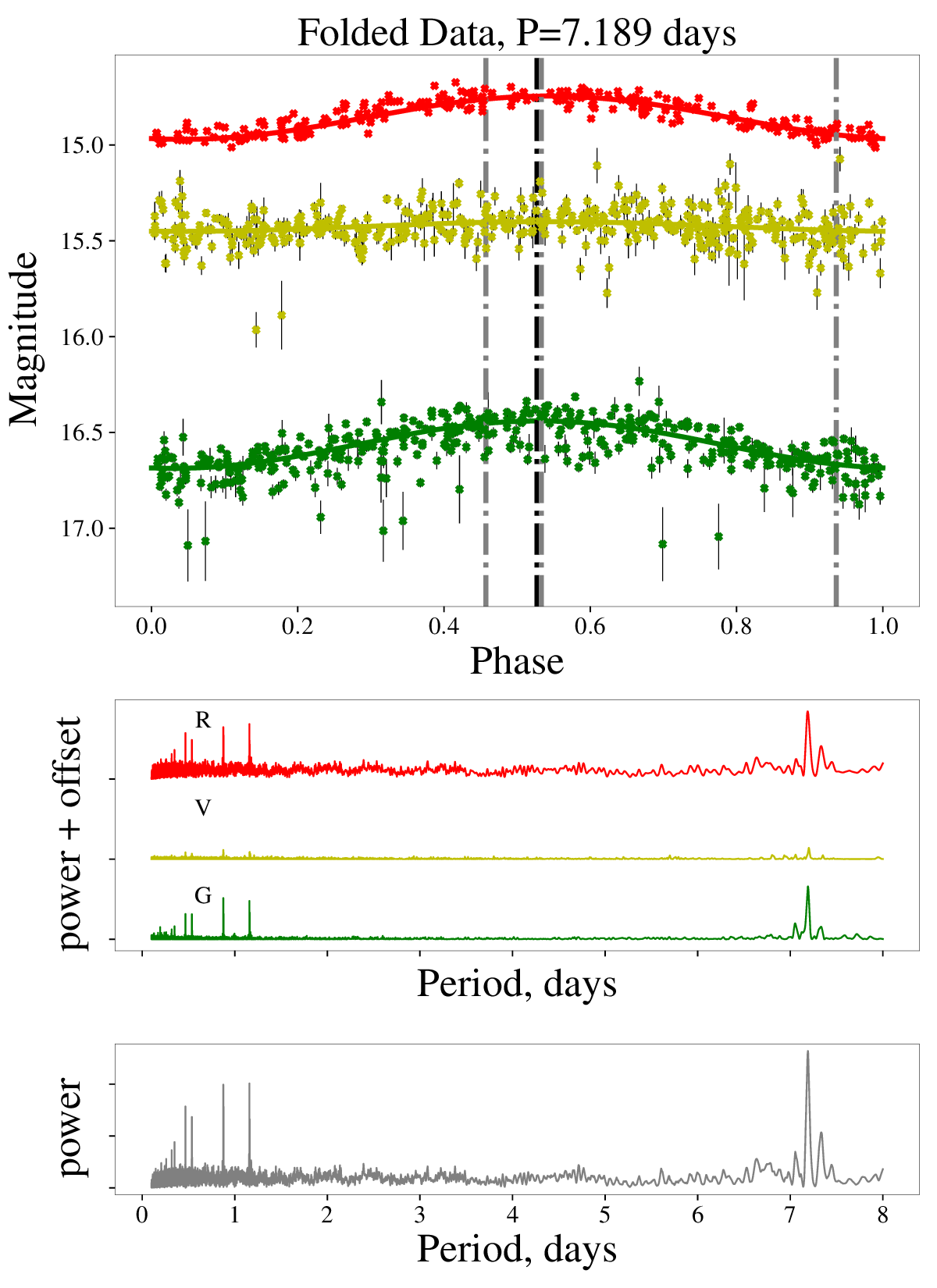}
\end{minipage}%
\begin{minipage}{.5\textwidth}
  \centering
  \includegraphics[width=1\linewidth]{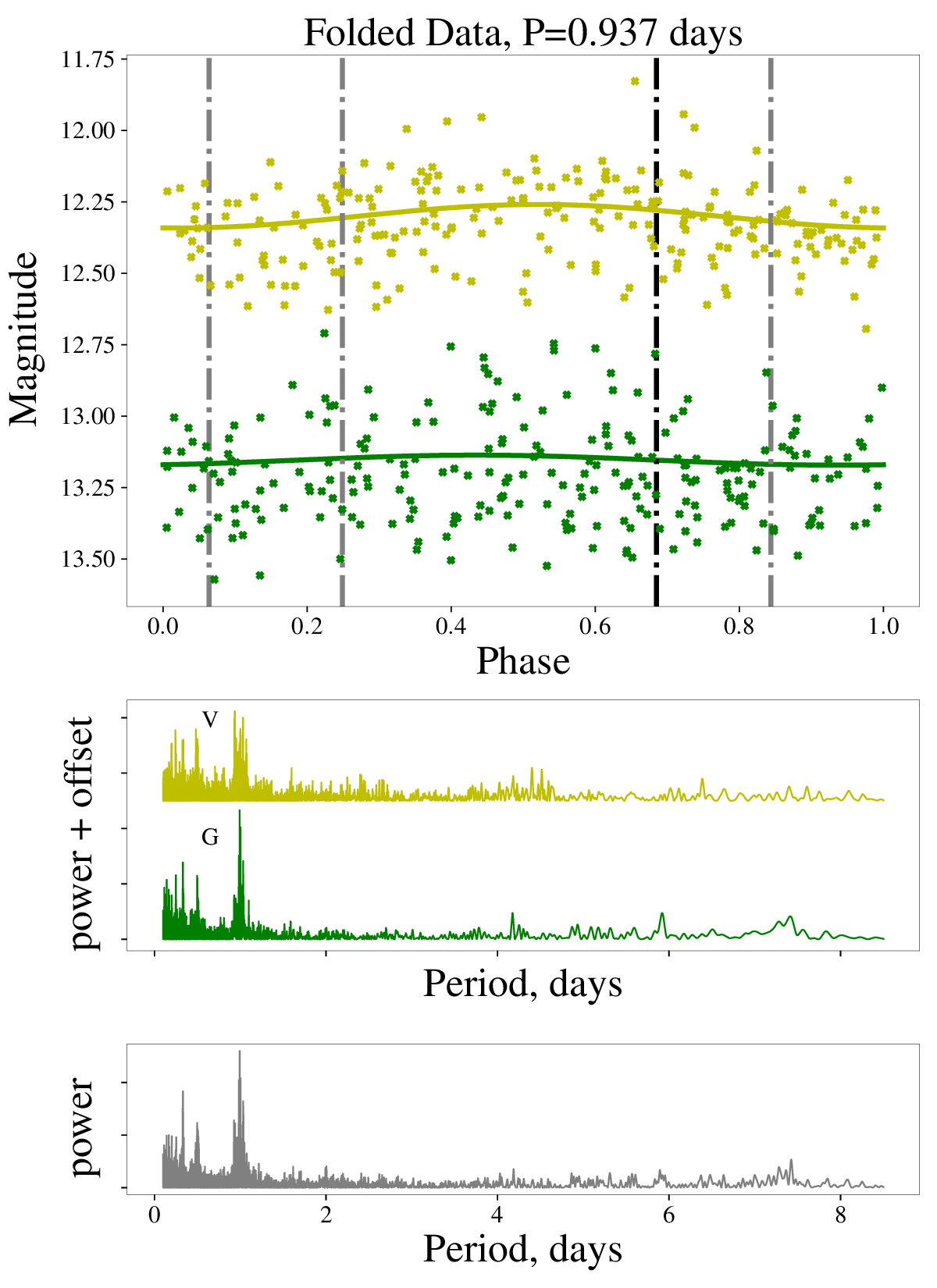}
\end{minipage}
\caption{Phase folded optical light curves of \xbIII{} (left image) \xbI{} (right image) from ZTF and ASAS-SN (top panels) with standard periodogram in each band (middle panels) and multiband periodogram~(bottom panels).}
\label{xrb:opt}
\end{figure}

\boldt{In order to examine the sources studied here for periodic behaviour in optical bands, we accessed archival data from the databases of ZTF, ASAS-SN and TESS. The corresponding light curves ($R$, $G$ and $V$ bands from ZTF and ASAS-SN data) are shown in Fig.~\ref{xrb:opt} for \xbIII{}~(left panel) and \xbI{}~(right panel). We conducted epoch folding on the light curves and produced periodograms using the python package \texttt{gatspy}\footnote{\url{https://github.com/astroML/gatspy}}, which is based on the classical Lomb-Scargle algorithm~\citep[][]{1976Ap&SS..39..447L,1982ApJ...263..835S}, see \cite{Gatspy} for more details. For both candidates, black and grey vertical lines indicate the observation times of the corresponding eRASS surveys (eRASS1 and eRASS2--4  respectively). We searched for periodic behaviour up to 300 days for both sources, but show only a part of the periodogram in Fig.~\ref{xrb:opt} for clarity.}

\boldt{In the case of \xbIII{}, the ZTF data allowed detection of some significant modes in the periodogram~(see lower and middle panels on the left image of Fig.~\ref{xrb:opt}). The two highest peaks on the periodogram with high precision are the same as those reported by ATLAS and ZTF (see Sect.~\ref{x_ray_pos}), at $1.158$ and $7.189$ days. However, all the peaks in the periodogram below ${\sim}7$ days with high accuracy are simply just beat tones (carrier and envelope frequencies with corresponding aliasing multiplicities) caused by the periodicity of $7.189$ days attributed to \xbIII{} and the Earth's rotation harmonic.}

\boldt{The right panel of Fig.~\ref{xrb:opt} shows the optical light curves (V and G band) of \xbI{}'s counterpart provided by ASAS-SN telescope (this source is out of sight of ZTF observatory). The peak signal detected by us differs from the one given in ASAS-SN~(see Sect.~\ref{x_ray_pos}), but the periodogram overall does not look very promising (see lower and middle panels on the right panel of Fig.~\ref{xrb:opt}).}

\begin{figure}[t]
\centering
\includegraphics[width=0.65\textwidth]{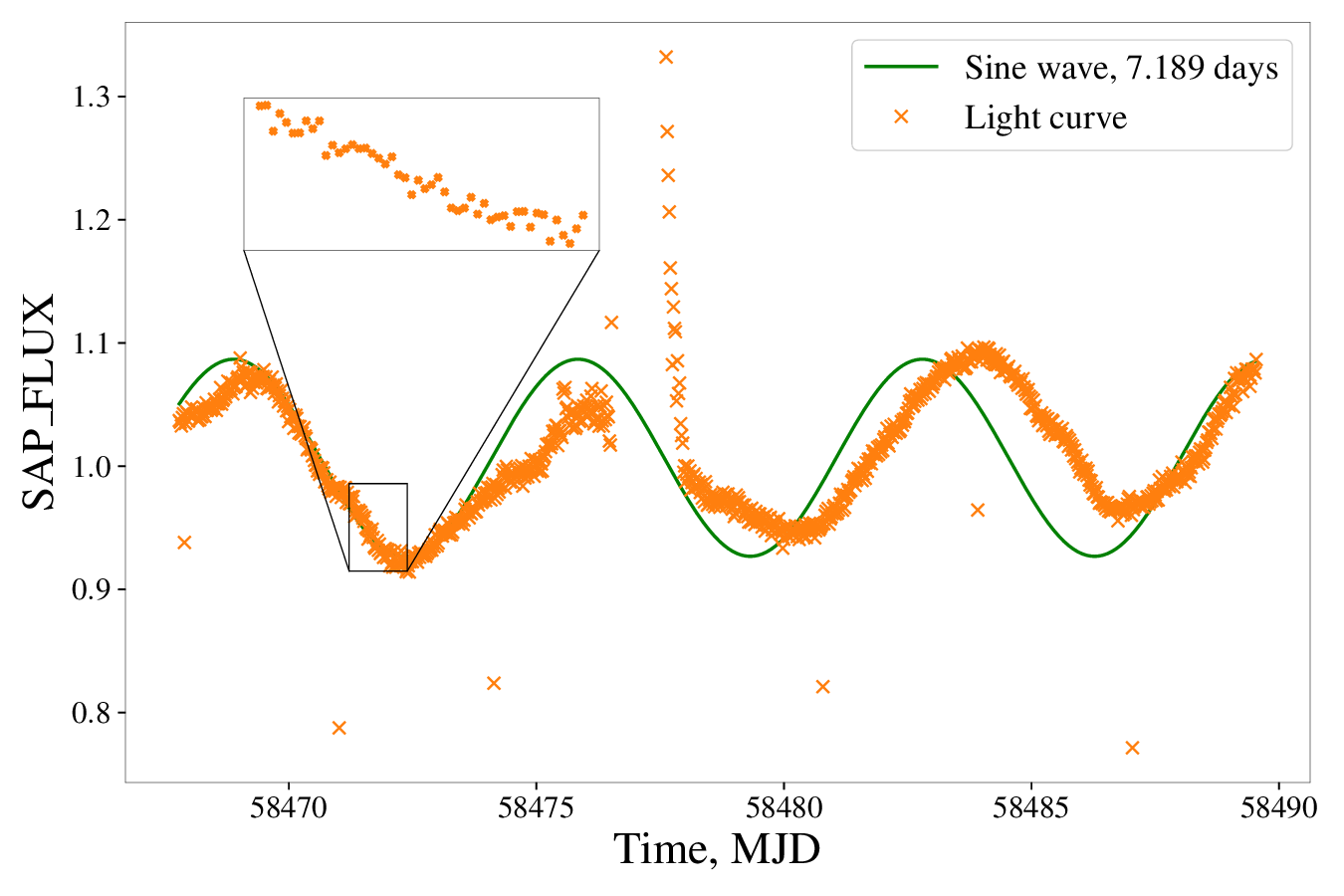}
\caption{TESS's optical light curve ($\sim$ 21.7 days duration) of the \xbIII{} counterpart along with the sine wave having a period of 7.189 days obtained from ZTF Lomb-Scargle periodogram. Observation is taken in 2018.}
\label{TESS_xb3}
\end{figure}

\begin{figure}[t]
\centering
\includegraphics[width=1\textwidth]{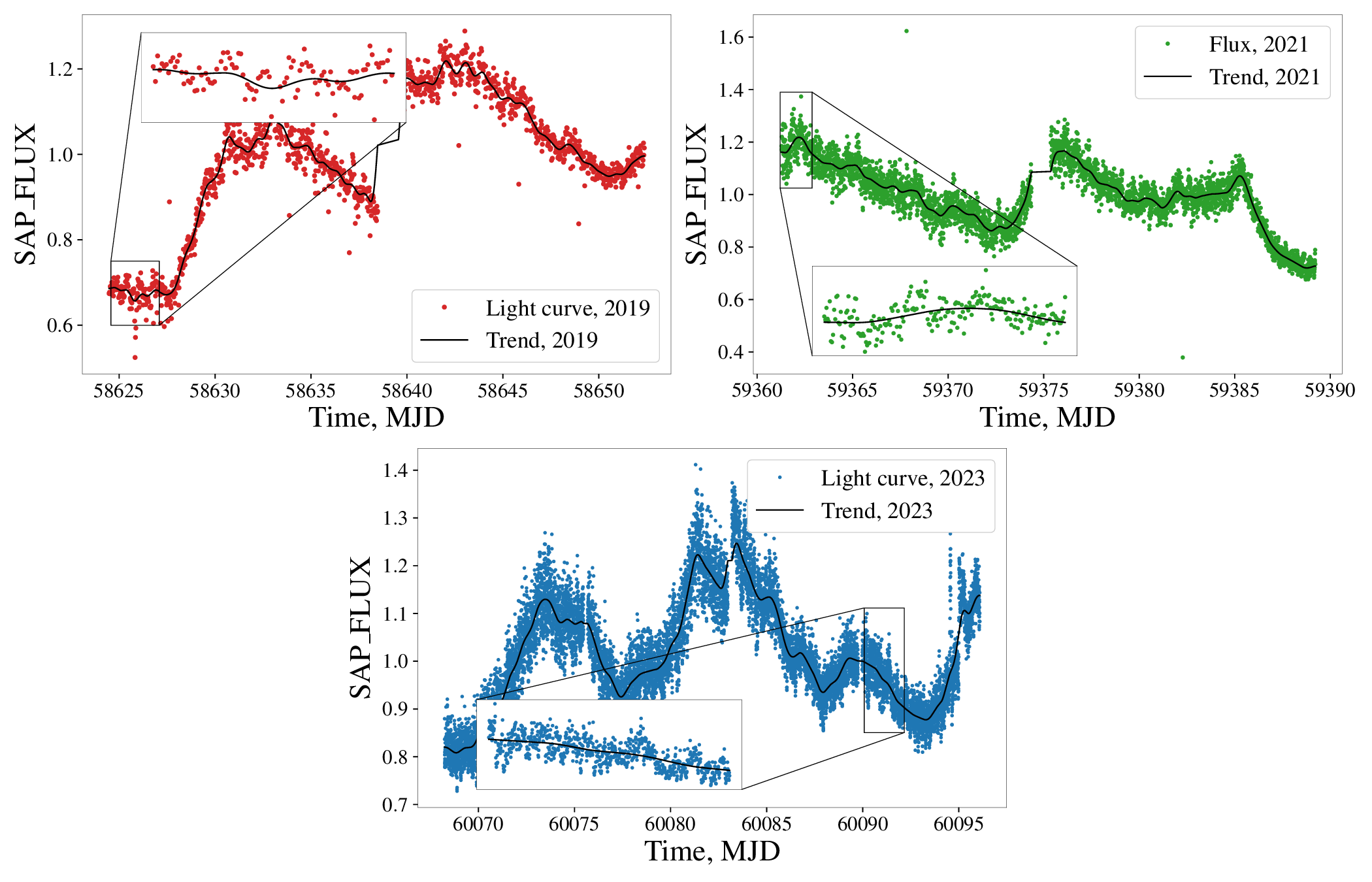}
\caption{TESS's optical light curves ($\sim$ 27.9 days duration each) of the \xbI{} counterpart. Black lines represent a trend catapulted by S-G smoothing filter. Left, right and bottom panels of the figure show observations taken in 2019, 2021 and 2023, respectively.}
\label{TESS_xb1}
\end{figure}

\boldt{Additionally, we looked for archival TESS lightcurves of the two sources, using the Mikulski Archive for Space Telescopes (MAST)\footnote{\url{https://mast.stsci.edu/portal/Mashup/Clients/Mast/Portal.html}}. The candidates only appear in Quick-Look Pipeline (QLP) TESS data, which means that we only have access to Simple Aperture Photometry (SAP) flux instead of also having Pre-search Data Conditioning SAP (PDCSAP) flux. PDCSAP flux is generally cleaner than the SAP flux and has fewer systematic trends.
Nevertheless, the corresponding light curves are presented in the Fig.~\ref{TESS_xb3} and Fig.~\ref{TESS_xb1} for \xbIII{} and 
\xbI{} respectively.}

\boldt{The light curve of \xbIII{}'s optical counterpart clearly shows a sinusoidal periodic signal, which is really close to the one found in ZTF data, namely 7.189\,days (Fig.~\ref{TESS_xb3}, green line). However, around 58475-80\,days, the observed light curve diverges from the sine wave, presumably due to a flare seen in that interval, after which, it resumes normal sinusoidal behaviour. Such short-term flares are common for late-type (usually M-type and later) chromospherically/magnetically active stars (especially for rich coronal X-ray emitters), and they are generally associated with stellar coronal mass ejection (CME) events~\citep[see][and references therein]{1984ApJ...279..763N,1989SoPh..121..299P,2015MNRAS.447.2714B,2017ApJS..232...26V}. Furthermore, we don't detect any light curve variations on a day/sub-day level~(Fig.~\ref{TESS_xb3}, see zoomed-in random section), which appear on the periodogram~(see Fig.~\ref{xrb:opt}, left image). This supports the idea that any periodic signals found below $\sim 7$ days are indeed just beat tones and their aliased versions. We can also see in the Fig.~\ref{xrb:opt} that the shape of the TESS light curve changes slightly ($\approx 0.1$\,mag amplitude) and is strongly affected by the flare. Thus, it seems that the 7.189\,day signal could be caused by starspots~\citep{2005LRSP....2....8B,2017A&A...599A...1S,2020A&A...644A..26I}, rather than come from the modulation of the orbital period.}

\boldt{The case of \xbI{} stands out in several respects. Figure~\ref{TESS_xb1} presents TESS observations of the source from 2019, 2021, and 2023. The 2023 epoch benefits from a significantly higher cadence --- $\sim$3\,min compared to $\sim$10\,min and $\sim$30\,min in 2021 and 2019 datasets, respectively, allowing for much finer time resolution. Each light curve displays distinct quasi/a-periodic variability on timescales of several days, with brightness fluctuations reaching nearly 1\,mag, consistent with trends seen in the ASAS-SN data.}

\boldt{Given the complex and high variable nature of \xbI{}'s photometric behaviour, we applied a de-trending procedure to the TESS light curves before searching for a periodic signal. As SALT spectroscopy identifies \xbI{} as a CV~(see Sect.~\ref{spec_xb1}), an orbital period on the order of a few hours is expected. To extract variations on this shorter timescale, we used a Savitzky-Golay filter~\citep[S-G,][]{1964AnaCh..36.1627S,2011ISPM...28..111S} to subtract long-term trends and variability exceeding 1\,day~(see removed trends in Fig.~\ref{TESS_xb1}, black lines).}

\begin{figure}[t]
\centering
\includegraphics[width=1\textwidth]{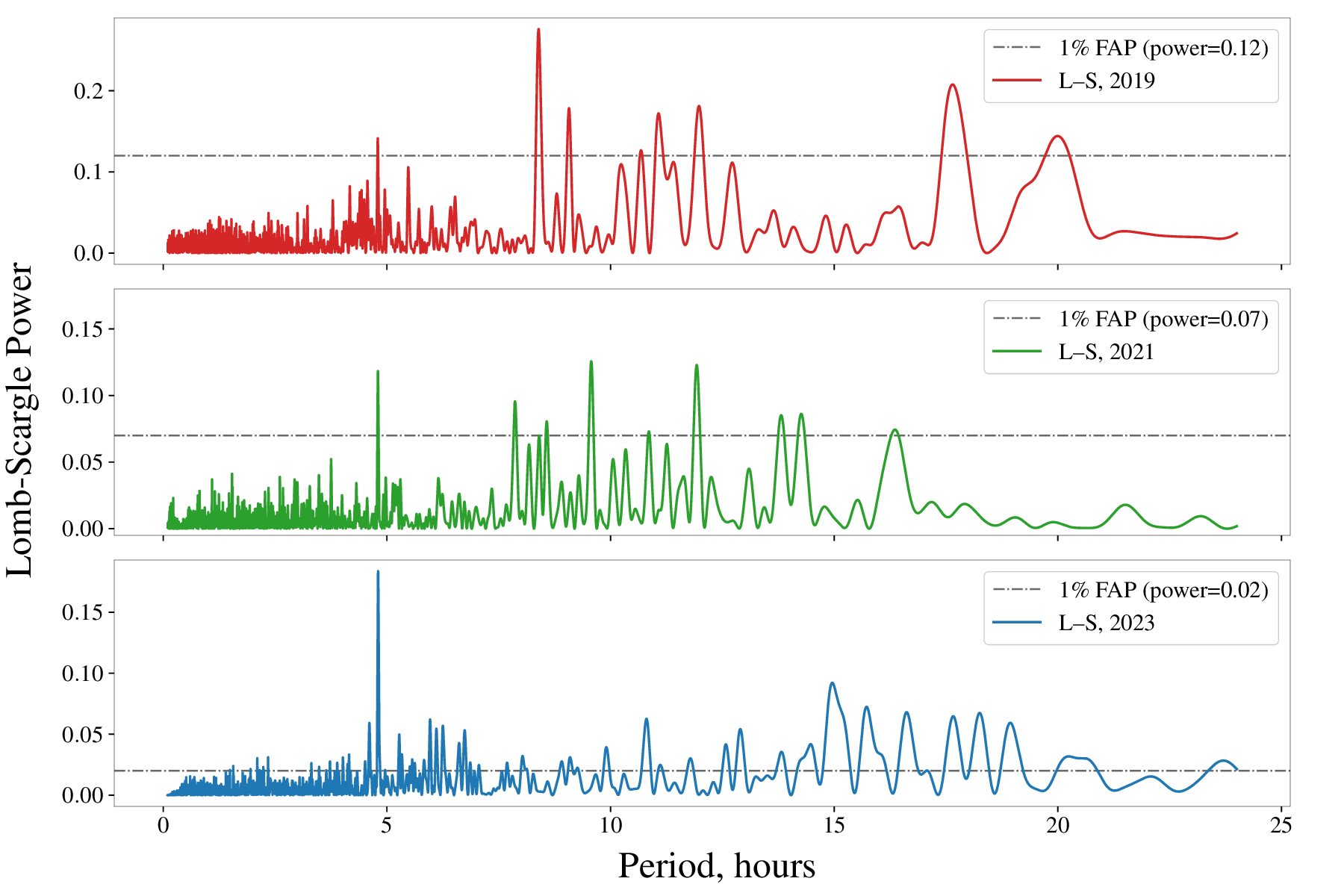}
\caption{Lomb-Scargle periodograms for \xbI{} based on de-trended TESS data from 2019, 2021 and 2023 observations (top, middle and bottom  panels, respectively). All three periodograms show a peak around 4.802\,h, which becomes more dominant with better cadence. Grey dash-dotted lines show the periodogram levels corresponding to a maximum peak false-alarm probability of 1\%.}
\label{TESS_period}
\end{figure}

\boldt{After de-trending, a Lomb-Scargle periodogram was computed for each TESS dataset~(see Fig.~\ref{TESS_period}). All three light curves, despite being collected over a span of four years, consistently show a strong, stable peak at around 4.802\,hours. While other peaks (seen in 2019 and 2021) are occasionally present, they seem to be transient and disappear in the better-cadence 2023 data. When combining all three epochs to bridge gaps between sectors without cycle count alias, the 4.802-hour peak remains to be one and only prominent feature. To determine the statistical significance of periodic signals, we employed a bootstrap resampling technique. This method generates synthetic datasets that preserve the original observational cadence, allowing to approximate the true distribution of peak maxima in the absence of true periodic variability. Through this computational approach, we established a detection threshold corresponding to a 1\% false-alarm probability for peak maxima, as illustrated in Fig.~\ref{TESS_period}.}

\boldt{To ensure that the choice of de-trending method did not artificially introduce or suppress any periodic signals, we tested several alternative smoothing techniques. These included moving average, modified sinc kernel, Whittaker–Henderson smoothing~\citep{Whittaker_1922,henderson1924graduation} and S-G filter with fitting weights~\citep[see][for the review]{SchmidSG}. In all cases, the same dominant periodic signal at $\sim$4.8\,hours was recovered, confirming the robustness of the result. We also scanned periodogram space up to several days by increasing the long-term cut-off in the de-trending process; however, no other significant peaks were found. 
Ultimately, the presence of such stable signal in all 3 observations spanned across 4 years strongly supports the identification of \xbI{}'s orbital period as 4.8\,hours.}

\subsection{Stellar parameters of \xbIII{} from SED}\label{SEDs}

\boldt{In order to determine the radii, effective temperatures, luminosities, and other stellar parameters of \xbIII{},
we performed a fit of the spectral energy distribution (SED) using spectrAl eneRgy dIstribution
bAyesian moDel averagiNg fittEr~\citep[{\sc ariadne},][]{2022MNRAS.513.2719V}\footnote{\url{https://github.com/jvines/astroARIADNE}}. This code is meant to fit MWL photometry to a different stellar atmosphere models based on nested sampling algorithms~\citep{Skilling2004,Skilling2006} within Bayesian inference framework, similar to the classic Markov Chain
Monte Carlo (MCMC) methods.}

\begin{figure}[t]
\centering
\includegraphics[width=0.97\linewidth]{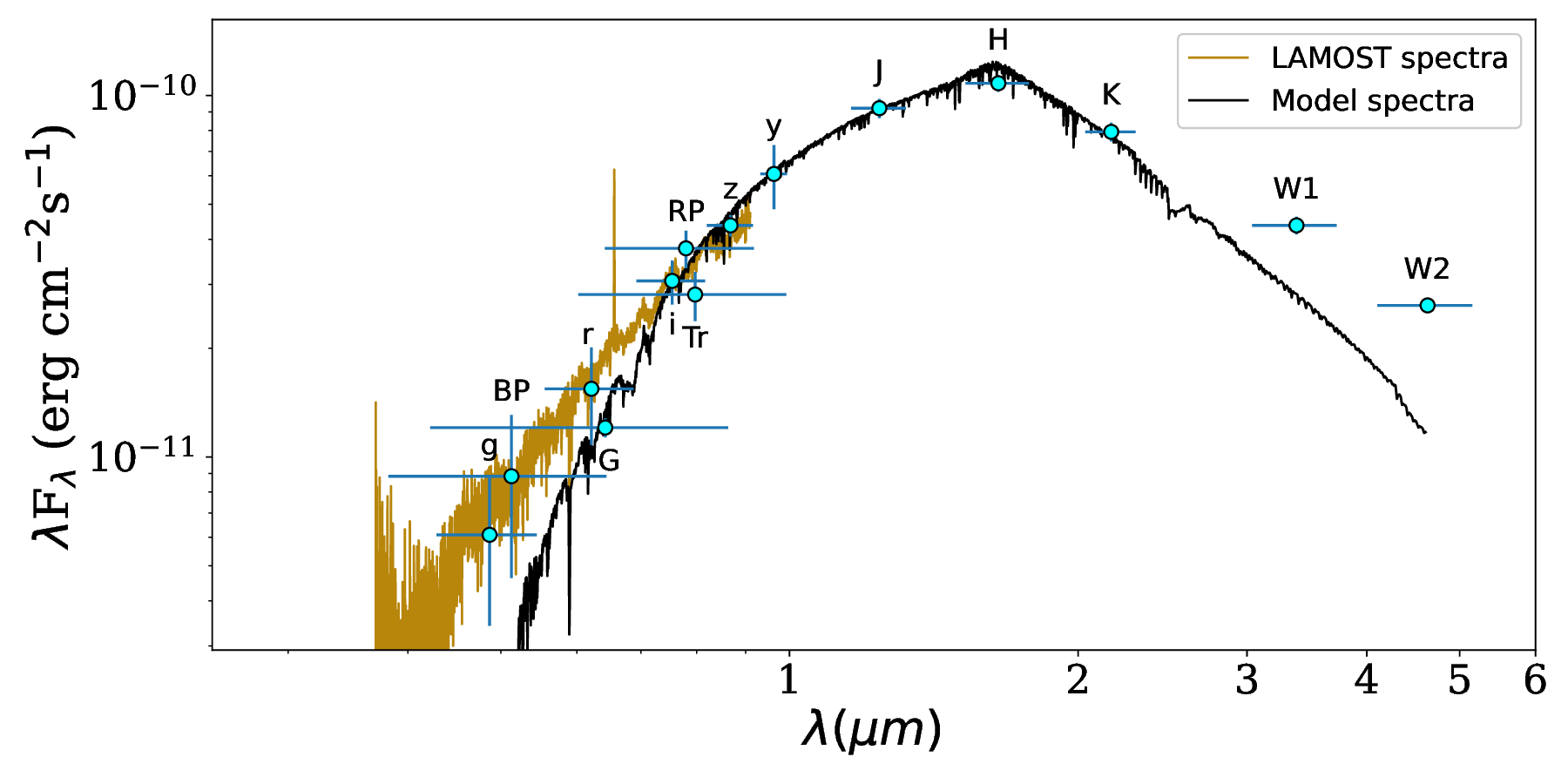}
\caption{Fit to the SED of \xbIII{}. LAMOST spectra (same as in the Fig.~\ref{xrb3_optspectra}), normalized to $i$-band magnitude, is shown in dark goldenrod colour. Model Phoenix flux is drawn with black, while blue errorbar dots represent corresponding photometry.}
\label{SED_fitJ06}
\end{figure}

\boldt{Corresponding best fit is shown in Fig.~\ref{SED_fitJ06}. The resulting best fit parameters are presented in Table~\ref{tab:ariadne}. SED fits were made using \texttt{dynesty} engine~\citep{Speagle2019} and Phoenix models~\citep{Husser2013}\footnote{\url{https://phoenix.astro.physik.uni-goettingen.de/data/HiResFITS/PHOENIX-ACES-AGSS-COND-2011}}, since they provided more accurate results based on the given photometry. The blue dots in Fig.~\ref{SED_fitJ06} indicate corresponding filter fluxes converted from the observed magnitudes. {\sc ariadne} uses specific set of allowed photometry filters\footnote{\url{https://github.com/jvines/astroARIADNE/blob/master/filters.md}}, therefore, we obtained every archival magnitude possible. 
Specifically, $g$, $r$, $i$, $z$, $y$ data for \xbIII{} were taken from the Panoramic Survey Telescope and Rapid Response
System (Pan-STARRS) catalogue~\citep{2002SPIE.4836..154K,2016arXiv161205560C}\footnote{\url{https://cdsarc.cds.unistra.fr/viz-bin/cat/II/349}}. In turn, $G$, $BP$, $RP$ information, $J$, $H$, $K$ data, as well as $W1$, $W2$ magnitudes/fluxes were taken from \textit{Gaia}\,DR3, 2MASS and WISE catalogues, respectively. TESS mean magnitude is presented as $Tr$.} 

\boldt{As can be seen in Fig.~\ref{SED_fitJ06}, the model predicts observations very well for \xbIII{} from the mid-optical to NIR range, but we have some excess in the blue optical and WISE IR bands. However, there is another bright IR source nearby \xbIII{} (\textit{Gaia}\,DR3 3345882607097252480, at an angular separation ${\sim}7\arcsec$), therefore, the observed excess in W1 and W2 WISE bands could be just the result of a blend. Another possibility is that the IR excess seen is due to gas surrounding the subgiant star, which is caused by its high surface mass-loss rate. Such behaviour is known among subgiant and AGB stars~\citep{2000ApJ...531..826C, 2010ApJ...711L..99B}. In turn, we believe that blue optical excess is small to be caused by the other binary component, and is likely just a result of insufficient absorption constraints~(see Table~\ref{tab:ariadne}) and an active chromosphere. Additionally, the resulting SED fit parameters on temperature, surface gravity and metallicity for \xbIII{} seem to be in line with LAMOST archival data~(see Tables~\ref{tab:lam} and \ref{tab:ariadne}).}

\boldt{We note that SED modelling of \xbI{}, classified as a CV~(see Sect.~\ref{spec_xb1} and upcoming discussion Sect.~\ref{disc}), requires an accurate treatment of the individual components of the system --- primarily the accretion disc, the boundary layer, and the white dwarf. The contribution from the donor star is expected to be minimal. Proper modelling therefore involves computing synthetic atmosphere spectra under specific physical conditions for each component, which lies beyond the scope and volume of the present study. A dedicated analysis focusing on \xbI{}’s SED will be presented in a forthcoming paper.}

\begin{table}[t]
\renewcommand{\arraystretch}{1.4} 
\renewcommand{\tabcolsep}{1.8mm}   
\centering
\caption{Parameters of {\sc ariadne}'s SED fit results for \xbIII{} optical counterpart.}
\begin{tabular}{@{}llllllll@{}}
\toprule  
$T_{\rm eff}$, K &
$\log(g)$, cm/s$^2$ &
Fe/H, dex &
$D$, pc &
$R$, $R_{\odot}$ &
$A_{\rm V}$, mag &
$L$, $L_{\odot}$ &
Age, Gyr \\
\midrule
$3885.54^{+388}_{-355}$ &
$4.47^{+0.73}_{-0.58}$ &
$-0.54^{+0.33}_{-0.43}$ &
$787^{+101}_{-71}$ &
$5.07^{+0.94}_{-0.61}$ &
$3.32^{+0.62}_{-1.48}$ &
$5.88^{+3.81}_{-1.04}$ &
$11.9^{+2.4}_{-3.9}$  \\
\bottomrule  
\end{tabular}
\label{tab:ariadne}
\end{table}

\section{Discussion}\label{disc}

Both candidates, in general, have similar initial X-ray observational appearance. Namely, despite relatively high brightness in the X-rays they appear to exhibit X-ray spectra different from those of accretion-dominated XRBs. Indeed, based on the results of spectral analysis, we conclude that spectra of both objects are well approximated with absorbed two-temperature \texttt{apec} model. This allows the accurate determination of the intrinsic X-ray fluxes in 0.2-2.3\,keV energy range as shown in Fig.~\ref{Xray_lc}. Unfortunately, we do not find X-ray pulsations for either candidate, which would help better constrain the nature of the sources based on X-ray data. 

\boldt{Spectral analysis and SED fitting results of \xbIII{} indicate the presence of a late M0-M2 type star~($T_{\rm eff} \lesssim 4000$\,K). Given the resulting radius and bolometric luminosity of this star ($R \sim 5 R_{\odot}$ and $L \sim 6 L_{\odot}$, see Table~\ref{tab:ariadne}) we identify  \xbIII{}'s optical counterpart as a subgiant. 
The source also shows a really strong H$_{\alpha}$ emission signature in its optical LAMOST spectra. However, the absence of other emission line features (such as $\mathrm{H_{\beta}}$, $\mathrm{H_{\gamma}}$, and He lines) typical for accreting binaries makes it unlikely for \xbIII{} to be an XRB or a CV. Therefore, H$_{\alpha}$ emission could be just a result of an active chromosphere. SED fit results also show that the source exhibits high reddening~($A_{\rm V} \sim 3.3 $\,) which is typical for stars at the suggested for \xbIII{}'s counterpart evolutionary stage and would be in line with high coronal activity~\citep{2006A&A...454..609M,2009AJ....138..615M,2011A&A...526A...4V}. }

\boldt{The optical counterpart of \xbIII{} is found to be periodically variable (the modulation is $7.189$ days). Additionally, we see a flare in its light curve~(see Fig.~\ref{TESS_xb1}), which is characteristic for late-type chromospherically active stars. The shape of its light curve suggests that the periodicity could be caused by starspot(s) rather than orbital modulation. Therefore, we can conclude that this candidate is most probably a M-type H$_{\alpha}$ emitting coronal active subgiant, which has surface starspot(s) causing $7.189$ days period found in its optical light curve.}

\boldt{Given the assumption that the 7.189\,days optical modulation of \xbIII{}'s counterpart is caused by starspot(s), with the stellar radius of $5.07^{+0.94}_{-0.61} R_{\odot}$~(Table~\ref{tab:ariadne}), we can calculate rotational velocity of the star to be $v_{\rm rot} = 35.7^{+6.6}_{-4.3}$\,km\,s$^{-1}$. However, the observed H$_{\alpha}$ emission profile~($\mathrm{FWHM_{H\alpha}} \approx 7.4 \mathrm{\r{A}}$) corresponds to a broad width of $\sim 330 $\,km\,s$^{-1}$, exceeding the expected rotational width by a factor of nine. Thus, assuming that H$_{\alpha}$ is bound with the star, FWHM suggests that some part of the emission may come from a region extending outward within a corotating circumstellar environment reaching up to 8-10 stellar radii. This has actually been seen and studied in other active stars~\citep{1981ApJ...251L.101R,1998A&A...338..191S,2004A&A...421..295Z,2019A&A...624A..83K}, where its been suggested that the overall H$_{\alpha}$ emission is composed of two parts: chromospheric and a circumstellar component.}

\boldt{In turn, \xbI{} has harder X-ray spectra compared to \xbIII{}, and more importantly, shows Fe line emission signatures around 6-7\,keV. However, in absence of accretion-dominance spectral features and the overall spectra resembling two temperature hot plasma, one can say that \xbI{} appears more as a CV or RS~CVn candidate rather than an XRB.}

\boldt{The SALT spectrum of \xbI{} reveals a blue continuum and prominent emission lines, including the H-Balmer series, He\,I, He\,II, and the C\,III/N\,III Bowen blend --- features characteristic of an accreting binary with the optically thin layer of hot plasma which goes in line with X-ray double apec spectrum~(see Fig.~\ref{Spec_all} and Table~\ref{tab:specparams}). The absence of donor star signatures in the SALT's spectrum and the overall spectral profile strongly support the classification of \xbI{} as a CV. Moreover, each Balmer emission line in \xbI{}'s optical spectrum is accompanied by the broad absorption wings, which are likely due to the linear Stark effect originating in an optically thick, high-gravity accretion disc viewed at a low inclination angle. The emission lines typically exhibit FWHM around 330-450\,km\,s$^{-1}$~(see Table~\ref{tab:SALT_em}), indicating the presence of rapidly moving material in an optically thin region, possibly situated above the disc.}

\boldt{All emission lines have EWs greater than -10\,\AA~($\mathrm{EW>-10} $\,\AA, or $\mathrm{|EW|<10} $\,\AA) with the He\,II\,4686 to the $\mathrm{H_{\beta}}$ ratio below 0.2. These spectral characteristics argue against \xbI{} being a magnetic CV, such as a polar or intermediate polar~\citep{1984AIPC..115...49V}. Ultimately, considering optical and X-ray spectral features of \xbI{}, its high optical brightness, constant detection in eRASS1-4~(Fig.~\ref{Xray_lc}), relatively low X-ray to optical flux ratio ($\log(F_{\rm x}/F_{\rm opt}) \approx -2.5$), \xbI{} is most likely a weakly magnetised, novalike CV. No dwarf nova outbursts were found in archival data, supporting this classification.
In such novalike systems the accretion disc's radiation dominates the majority of the optical and near IR SED~\citep{2010AIPC.1273..342B,2022MNRAS.510.3605I}. Therefore, it is not surprising that we don't see donor star's features in SALT spectrum of \xbI{}, since the accretion disc remains in a persistently hot, accreting state. Ultimately, resulting classification establishes \xbI{} as the second bright novalike CV found with eROSITA that is new to the literature~\citep[SRGt\,062340.2$-$265751 is the first, see][]{2022A&A...661A..42S}.}

\boldt{TESS light curves of \xbI{} from 2019, 2021, and 2023 exhibit strong quasi/a-periodic variability on multi-day timescales, with brightness changes up to 1\,mag. Such high optical variability is not so typical for novalike CVs~\citep[even though similar behaviour has been observed in other novalikes, see][]{2012NewA...17..570R,2023JBAA..133..115S} and may be the result of instabilities in the accretion flow within the disc around a WD~\citep[so called “stunted” outbursts, see][]{1998AJ....115.2527H,2011arXiv1105.1381B,2018AJ....155...61R}. However, their exact origin is still puzzling, as well as X-ray to optical correlation during such events, and necessitates simultaneous deep X-ray and optical observations. Nevertheless, a consistent, signal at 4.802\,hours was recovered using Lomb-Scargle methods on each TESS observation, which were de-trended from the quasi/a-periodic variability. The persistence of such signal over four years of TESS observations firmly establishes it as a tentative orbital period of \xbI{} binary. Nevertheless, we need time-resolved follow-up spectroscopy and high-speed photometry to unambiguously confirm the derived period, which will be done in the future.}

It is interesting, therefore, to also discuss the reasons why both objects (\xbI{} and \xbIII{}) were identified as XRB candidates. First of all, we note that the fluxes initially estimated in the preliminary analysis with the standard eROSITA pipeline, based exclusively on estimated count-rate, were significantly higher 
( $\sim 1\times10^{-12}$\,erg\,cm$^{-2}$\,s$^{-1}$ for \xbIII{} and
$\sim 1.9\times10^{-12}$\,erg\,cm$^{-2}$\,s$^{-1}$ for \xbI{}). This discrepancy is due the problems with earlier versions of the eROSITA analysis pipeline~\citep{2020SPIE11444E..4QD,2022A&A...661A..25H,2022A&A...661A...1B} and to an assumed spectral shape (which was not known prior to XMM-\textit{Newton} observations).
This can be seen in Fig.~\ref{Xray_lc} where it is apparent that for both sources the fluxes are by factor of 2-3 lower than what obtained in the early processing, and for \xbIII{} there is also a discrepancy by factor of two between the 'standard flux' estimate (assuming count rate to flux conversion factor tailored to AGNs), and flux determined directly from the spectrum. On the other hand, the agreement between eROSITA and XMM-\textit{Newton} derived fluxes for \xbIII{} for which the observations occurred close in time is excellent. 
It can be also noted that both sources appear to be variable, which is clearly also part of the rather different fluxes measured by eROSITA and XMM-\textit{Newton}. The observed flux variability is also accompanied by spectral variability, which can be illustrated by comparing the eROSITA and XMM-\textit{Newton} spectra (see Fig.~\ref{Spec_all}).
Indeed, as it can be seen in the lower panels of the figures, eROSITA residuals appear to deviate from the best-fit model mostly defined by XMM-\textit{Newton} data, even though, along with \texttt{const} we untied the \texttt{tbabs} column density parameter $N_{\rm H}$ in the fit (see Table~\ref{tab:specparams}).

This fact that both spectral shape and flux appear to be variable is quite relevant as the main parameters based on which the observed XRB candidates were selected were the broadband flux and hardness of the spectrum. The fact that the true flux was overestimated has also direct impact on the possible origin of the observed emission, as we discuss below. It should be also noted, that the observed flux of \xbI{} during the XMM-\textit{Newton} observation was by factor of five lower than that detected by eROSITA in each of the four surveys, and the observed spectrum during this time was harder than the one of \xbIII{} (see Table~\ref{tab:specparams}), which could be also be relevant for the final classification. We conclude, therefore, that the reason for mis-classification of both candidates was mainly the over-estimated X-ray flux and hardness in our preliminary analysis of eRASS1 data, which were amplified by the intrinsic variability of both objects.

\section{Conclusion}\label{conc}

In this paper, we reported on follow-up of the two candidates for potential XRB systems, \xbIII{} and \xbI{}, both selected based on the preliminary analysis of X-ray properties of sources in the first eROSITA X-ray survey (eRASS1) with XMM-\textit{Newton}. With XMM-\textit{Newton} observations we were able to refine X-ray positions and unambiguously identify the optical counterparts for both X-ray sources. We have carried out a detailed X-ray spectral and timing analysis for both candidates. 
In both cases it was found that the X-ray spectrum is more representative of an optically thin thermal plasma rather than of an XRB.

\boldt{Based on the LAMOST optical spectrum of \xbIII{} and the {\sc ariadne} SED fit results we were able to identify this candidate as a coronal active M-type subgiant. 
We also reported the detection of a strong H$_{\alpha}$ emission line and absorption features in the optical spectrum, which favour chromospherically active subgiant origin for this source. Furthermore, it was found that \xbIII{} is photometrically variable with a period of 7.189\,days. Based on the optical data analysis, we concluded that this periodicity is likely to come from starspot(s) on the subgiant's surface. TESS light curve also shows that this source has undergone a flare typical for coronal active X-ray emitting late-type stars. With the stellar radius obtained based on the {\sc ariadne} SED fit results, we report that the star is rotating at the velocity $v_{\rm rot} = 35.7^{+6.6}_{-4.3}$~km\,s$^{-1}$ (assuming that optical period is due to starspots).}

\boldt{\xbI{} exhibits harder X-ray spectra than \xbIII{}, with iron line emission around 6–7 keV, but lacks accretion-dominated features. Its SALT optical spectrum shows a blue continuum, H-Balmer, He\,I,II and Bowen emission lines, as well as broad H-Balmer absorption wing profiles coming from an optically thick accretion disc around a WD. Emission line features disfavour magnetic CV origin scenario. Coupled with high optical brightness, persistent eRASS1-4 detection, 
colours, and absence of detected dwarf nova outbursts, \xbI{} is classified as a weakly magnetised novalike CV. We also tentatively identify the orbital period for \xbI{} as 4.802\,hours based on TESS observations, pending time-resolved spectroscopic confirmation. The identification of \xbI{} as a novalike CV marks only the second instance of such a system uncovered by eROSITA, following SRGt\,062340.2$-$265751~\citep{2022A&A...661A..42S}. 
This milestone underscores eROSITA’s expanding role in revealing more accreting binaries of different subclasses. In particular, pairing eROSITA’s X-ray surveys 
with large-scale spectroscopic follow-up 
programs --- such as eROSITA/SDSS 
\citep{2017K} --- will continue to 
advance the Galactic census of XRBs and CVs.}
 
The main reason for the selection of \xbIII{} and \xbI{} as XRB candidates was their relatively high X-ray flux in eRASS1 estimated at the time of selection using the preliminary eROSITA processing pipeline (c946) available at the time.
Both the analysis of XMM-\textit{Newton} data and the re-processing of eROSITA data with the current pipeline (eRASS1, c020) results in a significantly lower flux, and both objects would now not be selected as plausible XRB candidates even when using eRASS1 data only. Further refinement of classification pipeline is also now possible when considering data from eRASS1-4. We emphasise, however, that variability implies also that candidate selection shall still be done based on detections in individual surveys as most of the XRBs are intrinsically variable and thus non-detection or low observed X-ray flux in one survey does necessarily imply that an object is not a plausible XRB candidate. 

Nevertheless, considering the fact that the two other XRB candidates selected for follow-up with NuSTAR based on eROSITA data were in the end classified as XRBs~\citep[see][]{Doroshenko22Puls,2024arXiv241102655Z}, we conclude that the success rate for XRB candidates selection based on eRASS1 data only and simplified selection criteria already is significant.
In other words, we have already demonstrated the ability of eROSITA to identify new XRBs even if there is room for improving the accuracy. 
We are working on an improved candidate identification pipeline taking into account these conclusions, and plan to release the final catalogue of eROSITA XRB candidates in a separate publication accompanying the first public data release.

\bmhead{Acknowledgements}\label{thanx}
This work is based on data from eROSITA, the primary instrument aboard SRG, in the interests of the Deutsches Zentrum für Luft- und Raumfahrt (DLR). The SRG spacecraft was built by Lavochkin Association (NPOL) and its subcontractors, and is operated by NPOL with support from the Max Planck Institute for Extraterrestrial Physics (MPE). The development and construction of the eROSITA X-ray instrument was led by MPE, with contributions from the Dr. Karl Remeis Observatory Bamberg \& ECAP (FAU Erlangen-Nuernberg), the University of Hamburg Observatory, the Leibniz Institute for Astrophysics Potsdam (AIP), and the Institute for Astronomy and Astrophysics of the University of Tübingen, with the support of DLR and the Max Planck Society. The Argelander Institute for Astronomy of the University of Bonn and the Ludwig Maximilians Universität Munich also participated in the science preparation for eROSITA. The eROSITA data shown here were processed using the eSASS/NRTA software system developed by the German eROSITA consortium. We thank the XMM-\textit{Newton} Science Operations Centre (SOC) for their invaluable help in the quick scheduling of the observations. Some of the observations reported in this paper were obtained with the Southern African Large Telescope (SALT)\footnote{Under program 2024-2-LSP-001 (PI: David Buckley)}. This research has also made use of data, software and/or web tools obtained from the High Energy Astrophysics Science Archive Research Center (HEASARC), a service of the Astrophysics Science Division at NASA/GSFC and of the Smithsonian Astrophysical Observatory's High Energy Astrophysics Division. 
This research has made use of the VizieR and HEASARC database systems for querying objects and getting information from different catalogues. Optical photometric data were taken from TESS archive, ASAS-SN database, SNAD ZTF Viewer and \textit{Gaia}~DR3. SIMBAD database, operated at CDS, Strasbourg, France, was also used to get additional information.

\bmhead{Author contributions}\label{authcontr} All authors contributed to the study conception and design. A. A. conducted majority of the analysis and wrote most of the initial manuscript text. A. Z. conducted part of the eROSITA analysis, contributed to the manuscript text. V. D. is a PI of the XMM-Newton observations which were analysed in the paper, supervised the project and contributed to the manuscript text. A. Sch. and V. S. contributed to the analysis, manuscript's text and conclusions made. D. B. is a PI of the SALT observations analysed in the manuscript, obtained an identification spectrum. J. B. did the reduction the obtained SALT spectroscopy. J. W. and A. San. both supervised the project, and contributed to the manuscript text.

\bmhead{Funding}\label{fund} We thank Deutsche Forschungsgemeinschaft (DFG) within the eROSTEP research unit under DFG project number 414059771 for support (DO 2307/2-1 and SA 1777/4-2). 

\bmhead{Data availability}\label{data_avail} XMM-Newton\footnote{\url{http://nxsa.esac.esa.int/nxsa-web}} (Obs. Ids: 0883460301 and 0883460101) and eRASS1 data\footnote{\url{https://erosita.mpe.mpg.de/dr1/}} used in the study are publicly available. SALT spectral data are available upon request.

\bmhead{Competing interests}\label{int} The authors declare no competing interests.


\bibliography{sn-bibliography}

\end{document}